\documentstyle[aps,prb,epsf,amsmath]{revtex} 

\newcommand{\lp}{l^{+}}
\newcommand{\lm}{l^{-}}
\newcommand{\np}{n^{+}}
\newcommand{\nm}{n^{-}}
\newcommand{\mb}{\overline{m}}
\newcommand{\nb}{\overline{n}}
\newcommand{\npb}{\overline{n^{+}}}
\newcommand{\nmb}{\overline{n^{-}}}
\newcommand{\avg}[1]{\langle #1 \rangle}
\newcommand{\im}{{\rm Im}}
\newcommand{\dif}{{\rm d}}
\newcommand{\phit}{\tilde{\varphi}}

\newcommand{\dos}{DOS}
\newcommand{\oneD}{1D}

\begin{document}

\draft \title{Distribution function of the local density of states of
  a one--channel weakly disordered ring in an external magnetic field}
\author{H. Feldmann, E. P. Nakhmedov\cite{byline} and R. Oppermann}
\address{Institut f\"ur Theoretische Physik, Universit\"at W\"urzburg,
  97074 W\"urzburg, F.R.Germany} \date{\today}
\maketitle

\begin{abstract}
  A real space diagrammatic method, which is an extension of the
  Berezinskii technique to problems with periodic boundary condition,
  is formulated to study the density of states (\dos) $\rho(\epsilon,
  \phi)$ and its moments for a one--channel weakly disordered ring
  threaded by an external magnetic flux $\phi$. The exact result
  obtained for the average value of the \dos\ shows that $\rho(\epsilon,
  \phi)$ oscillates with a period of the flux quantum
  $\phi_0=\frac{hc}{e}$. However all higher moments of the \dos\
  oscillate with the halved period $\frac{\phi_0}{2}$. The exact
  expression for the \dos\ is valid for both weak localization ($L \gg
  l$, where $L$ is the rings circumference and $l$ is the mean free
  path) and ballistic ($L \leq l$) regimes. In the weak localization
  regime the distribution function of the \dos\ is calculated, which
  turns out to be of logarithmic normal form.
\end{abstract}

\pacs{
71.23.-k, % Electronic structure of disordered solids
71.23.An, % Theories and models; localized states
73.23.-b % Mesoscopic systems
}

\section{Introduction}
Interference effects in low--dimensional disordered conductors still
attract attention from both experimental and theoretical physicists,
although all main features and a lot of new effects have already been
discovered during the last twenty years. The prediction of the
oscillation in the kinetic coefficients in multiply connected
disordered normal metals in an external magnetic
field\cite{altshuler81a} and its experimental
observation\cite{sharvin81a} in a Mg cylinder was a very excellent
examination of weak localization phenomena, since the coherence of the
electron wavefunction during the circulation of a closed contour is
required to observe an oscillation. The period of oscillation of the
magnetoresistance predicted and observed firstly was equal to half of
the magnetic flux quantum $\phi_0=\frac{hc}{e}$. Further improvements
of the experiments on rings with large diameters and small widths gave
rise to the observation of a magnetoresistance oscillation with the
period $\phi_0$[\onlinecite{webb85a}], furthermore in the experiment of
Chandrasekhar {\it et al.}\cite{chandrasekhar85a} both periods were
observed.\cite{webb86a,aronov87a} Such a complex magnetic field
dependence of the magnetoresistance seems to be related with
the statistical properties of the sample.

Another development in theory was the prediction of a persistent
current in a one--channel disordered isolated loop.\cite{landauer83a}
Due to the analogy between loop and one--dimensional (\oneD) lattice with
a period equal to the circumference of the loop, a circulating current in
rings was suggested which is a periodic function of the enclosed flux
with a period of $\phi_0$.  Studies of the effects on the persistent
current at finite but small temperature and weak inelastic scattering
show that both weak inelastic and elastic scattering do not destroy
it.\cite{landauer85a,riedel88a,riedel89a,imry97a}  In experiments, the
persistent current was also
observed.\cite{levy90a,webb91a,mailly93a,mailly95a} A magnetization
measurement was performed in [\onlinecite{levy90a}] on $N=10^7$ disconnected
copper loops at $T<1.5\rm K$ where the electron phase coherence length
$L_\varphi$ exceeds the loop's circumference $L=2.2 {\rm \mu m}$, and it
shows evidence for a flux--periodic persistent current with halved
period and $3 \times 10^{-3}\frac{ev_F}{L}$ amplitude per ring, which is
remarkably higher than the theoretically expected value for the
persistent current per ring, $\sqrt{\frac{l}{L N}}\frac{ev_F}{L}$ [\onlinecite{riedel89a,montambaux90a,montambaux91a}].  Measurements of the persistent current
on single loops (with at least a few channels) in the
diffusive\cite{webb91a} ($L=8{\rm \mu m}$ and $l=70 {\rm nm}$, where $l$ is the
elastic mean free path) and ballistic\cite{mailly93a} ($L=8.5 {\rm \mu m}$
and $l=11{\rm \mu m}$) regimes reveal the period $\phi_0$. The amplitude of
the harmonic with $\frac{\phi_0}{2}$ was measured to be smaller by a
factor of 2-3 than that of the $\phi_0$ harmonic in [\onlinecite{webb91a}].

Effects of impurities in most of the theoretical investigations were
taken into account by the transfer matrix method according to the
Landauer expression and by generalizations of this method to
$n$--channel systems.\cite{imry86a,landauer85b,cheng97a,imry97a}
Although the Landauer formula gives full--flux periodicity for all
physical parameters, averaging over ensembles of
rings\cite{riedel88a,montambaux90a,montambaux91a,montambaux89a,riedel91a,altshuler91a,schmid91a}
or the calculation of the dynamical current instead of the
thermodynamical
potential\cite{efetov91a,efetov92a,kopietz92a,kopietz95a} were
suggested as an explanation of the observed halved periodicity.  In
the process of averaging over different impurity realizations in the
ensemble, the number of particles in each ring is proposed to be
constant, i.e.\ the persistent current is assumed to be determined by
the thermodynamic potential instead of the grand canonical
potential.\cite{riedel91a,altshuler91a,schmid91a}  Although there has
been done a lot of work on the Aharonov--Bohm effect, the existing
theories of non--interacting electrons still can explain neither the
high value of the experimentally observed persistent
currents\cite{levy90a,webb91a,mailly93a,mailly95a} nor its diamagnetic
sign.\cite{mailly95a} This can be partially connected with the
complexity of the experiments, particularly with difficulties of the
separation of phase effects in the rings with relatively large width
(larger than the mean free path)\cite{levy90a,mailly95a} from orbital
ones.

On the other hand, correlation effects may be a reason for the
discrepancy between theory and
experiment.\cite{weidenmueller98a,eckern90a,weidenmueller93a,weidenmueller94a,abraham93a,montambaux94a,kato94a,giamarchi94a,bouchiat95a,jagla93a}
Unfortunately, there is still no agreement on the effects of Coulomb interaction on the amplitude of the persistent current. Studies based on spinless electrons
in \oneD\ 
continuum\cite{weidenmueller98a,weidenmueller93a,weidenmueller94a} and
lattice models\cite{abraham93a,montambaux94a,kato94a} gave
controversial results, so the amplitude of the persistent current was
shown to be increased up to its disorder--free value according to the
former model, but a Mott--Hubbard metal--insulator transition in the
latter model was found to reduce the amplitude, or at least a possible
increase of the amplitude was negligibly small. There was also the
suggestion that correlation can change the fundamental period with the
magnetic flux and create fractional periodicity in a \oneD\
ring.\cite{jagla93a} Impurities and correlation acting together are again
subject of controversy. Thus, considering weak localization
corrections in first order in the electron--electron interaction to
the grand canonical potential,\cite{eckern90a} a persistent current
with a period of $\frac{\phi_0}{2}$ and an amplitude of $\sim \frac{e
  v_F l}{L^2}$, corresponding to the experiment,\cite{levy90a} was
obtained, while Monte Carlo simulation on a \oneD\ Luttinger
liquid\cite{mori95a} resulted in a persistent current with period
$\phi_0$ and with an amplitude decreased through interaction.
Resumming the existing results it can be said that neither the
noninteracting electron model nor models of correlated electrons yet
gave satisfactory answers to the questions put forward by the
experiments. These concern the period (under what condition both
periods or only the halved period are observed), the amplitude of the
persistent current and its diamagnetic sign (for the correlated
electron model the diamagnetic sign requires an attractive interaction
between the electrons).

To prevent the interference of the orbital effects in the presence of
an external magnetic field with the phase effects, the width of the
ring should be chosen as narrow as possible, i.e.\ a one--channel ring
with random impurities seems to be an ideal tool to study the
Aharonov--Bohm effect. However, interference effects in \oneD\ disordered
systems as a result of the coherent backscattering processes are
strong\cite{mott61a} irrespective of the degree of randomness. The
diffusion approximation, which was used in previous studies, is not
acceptable in \oneD\ systems even for the case of weak disorder.

Approaching the problem thoroughly, we use in this paper a weakly
disordered noninteracting electron model and construct for it a new
exactly solvable diagrammatic method, which is an extension of
Berezinskii's method\cite{berezinskii73a,gogolin82} to the problem
with periodic boundary conditions. Within this model, we sum up all
impurity scattering diagrams in the framework of the Born
approximation.

In Sec.\ref{sec:description}, we describe the method.  We calculate
the density of electronic states (\dos) in Sec.\ref{sec:dos}. Indeed an
average value is not enough to describe the observable parameters in
low dimensional mesoscopic systems. It is well known that the physical
parameters of a mesoscopic system with size $L$ satisfying the
condition $l<L\ll L_\varphi$ fluctuate from sample to
sample, i.e.\ self--averaging is
violated.\cite{altshuler85b,stone85a,webb91b} At $T=0$ all sufficiently large
systems become mesoscopic. In this case, high moments give a considerable
contribution,\cite{wegner80b,altshuler86b,altshuler86a,melnikov81a,abrikosov81a,altshuler89a,nakhmedov90a}
which results in strong differences between average value and typical
one of the observable parameter, i.e.\ the average value loses its
significance to characterize the experimental observation.  In
Sec.\ref{sec:moments} the diagrammatical method is applied to find the
$k$th moments of the \dos, $\avg{\rho^k(\epsilon,\phi)}$. The obtained
equations for $\avg{\rho^k(\epsilon,\phi)}$ show that, in contrast to
the average value of the \dos, all higher moments oscillate with the
halved period, $\frac{\phi_0}{2}$.  Although the structure of the
equations is complicated, the latter can be solved for the weak
localization regime when the condition $l\ll L$ is satisfied. This
procedure is given in Sec.\ref{sec:distribution}.

The zeroth (not depending on $\phi$) and the first (oscillating with
$\frac{\phi_0}{2}$) harmonic of $\avg{\rho^k(\epsilon,\phi)}$ are
studied explicitely in this section. Both are shown to increase with
$k$ as $\exp(k^2)$, which gives rise to a logarithmic normal
distribution.

%In the concluding section~\ref{sec:conclusion} we discuss the
%possibility to explain the high amplitude of the experimentally
%observed persistent current in the framework of our method due to the
%impurity induced Dyson singularity of the \dos\ in the middle of the
%band of a \oneD\ lattice model.

In section~\ref{sec:conclusion} we conclude our results and
discuss possibilities to extend our approach to related problems.
  
\section{Description of the method}
\label{sec:description}

We consider here a one--channel disordered ring with length $L=2 \pi
r$, threaded by a magnetic flux $\phi$ through the opening. The
Hamiltonian of the system can be written as
\begin{equation}
H= \frac{\hbar^2}{2 m^* r^2} \bigl( i \frac{\partial}{\partial \varphi} +
\frac{\phi}{\phi_0} \bigr)^2 + V_{\rm imp}(\varphi)
\label{eq:hamiltonian}
\end{equation}
where $\phi_0=\frac{hc}{e}$ is the fundamental period of a flux
quantum and $m^*$ is the effective mass of an electron.
The potential $V_{\rm imp}$ of randomly distributed
impurities is considered to be weak, so that scattering processes can
be studied in the framework of the Born approximation. Below, we
use the spatial variable $x=r \varphi$ instead of the angle $\varphi$.

Our aim in this section is to construct a diagrammatical method for
the calculation of the average values of the \dos,
$\avg{\rho(\epsilon,\phi;L)}$, and of its moments
$\avg{\rho^n(\epsilon,\phi;L)}$. The bracket $\avg{\dots}$ denotes the
average over the impurity configurations.  Expressing the \dos\ by means
of retarded ($G_R$) and advanced ($G_A$) Green's functions (GF) as
\begin{equation}
\rho(\epsilon,\phi;x)=-\frac{1}{\pi} \im G_R(\epsilon,\phi;x,x)
=\frac{1}{2 \pi i} [G_A(\epsilon,\phi;x,x)-G_R(\epsilon,\phi;x,x)]
\label{eq:dos_formula}
\end{equation}
the $n$--th moment of $\rho(\epsilon,\phi;x)$ can be given by
\begin{equation}
\avg{\rho^k(\epsilon,\phi;x)}=\frac{1}{(2 \pi i)^n} 
\sum_{l=0}^{k} \binom{n}{k} (-1)^l \avg{G_R^l(\epsilon,\phi;x,x)G_A^{k-l}
(\epsilon,\phi;x,x)}
\label{eq:rhok_definition}
\end{equation}

The Berezinskii diagram technique \cite{berezinskii73a,gogolin82} is
applied to calculate the average value of a single GF and
higher--order correlators $\avg{G_R^lG_A^{k-l}}$. In contrast to
strictly \oneD\ disordered wires, quantum corrections to the \dos\ of a ring
turn out to exist even for the weakly disordered limit due to
periodicity.

As for infinite systems, we consider as starting point a free particle
with wave function $\psi_p(x) \propto \exp(i p x)$, where the momentum
can assume arbitrary values, leading to a continuous spectrum
$\epsilon_p=\frac{\hbar^2}{2 m^*}(p-\frac{\phi}{\phi_0 r})^2$. The
``bare'' GF $G_{R,A}^0$ can be calculated easily:
\begin{equation}
G_{R,A}^0(\epsilon,\phi;x,x')=\int \frac{dp}{2 \pi} \frac{e^{ip(x-x')}}
{\epsilon-\epsilon_p \pm i \eta} 
= \mp \frac{i}{v(\epsilon) \hbar} e^{i 2 \pi \frac{\phi}{\phi_0}
\frac{x-x'}{L} \pm i p(\epsilon)|x-x'| - \frac{\eta}{v(\epsilon)}
|x-x'|}
\label{eq:bareGF}
\end{equation}
where $L$ is the circumference of the ring and the parameter $\eta$
is introduced phenomenologically to model inelastic processes, which 
result in a blurring of the energy levels.
$v(\epsilon)=\sqrt{\frac{2 \epsilon}{m^*}}$ and $\hbar p(\epsilon) =
\sqrt{2 m^* \epsilon}$ are the velocity and the momentum of an
electron with energy $\epsilon$, respectively.
Notice that Zeeman splitting has not been taken into account here. For an electron with spin $s=\pm \frac12$, it would throughout the paper lead to a shift 
of the energy as $\epsilon \to \epsilon - s g \mu_B B$ (where $g$ is the gyromagnetic ratio of the electron and $\mu_B$ is the Bohr magneton).

This bare GF, however, is not yet the real GF for an electron in a
ring without impurities, since it does not reflect the finite size of
the system and the periodic boundary conditions. These are taken into
account by allowing the particles to make arbitrary revolutions around
the ring, which leads to the expected quantization effect. According
to this prescription, the GF for a clean ring $\tilde{G}_{R,A}^0$ is
\begin{equation}
\tilde{G}_{R,A}^0(\epsilon,\phi;x,x')=
G_{R,A}^0(\epsilon,\phi;x,x')+G_{R,A}^0(\epsilon,\phi;x,x'+L)
+G_{R,A}^0(\epsilon,\phi;x,x'-L)+G_{R,A}^0(\epsilon,\phi;x,x'+2L) \dots
\label{eq:GF_clean_ring}
\end{equation}
One may verify this approach by calculating the \dos\ of a clean ring in
a magnetic field $\rho_0(\epsilon,\phi)$ from Eqs.(\ref{eq:dos_formula}), (\ref{eq:bareGF}), and (\ref{eq:GF_clean_ring}):
\begin{equation}
\rho_0(\epsilon,\phi)=\rho_0 + 2 \rho_0 \sum_{n=1}^{\infty} 
\cos(p(\epsilon) L n)
\cos(2 \pi \frac{\phi}{\phi_0}n) e^{-\frac{\eta}{v(\epsilon)}Ln}
\label{eq:pureDoS}
\end{equation}
where $\rho_0=1/(\pi \hbar v(\epsilon))$ is the \dos\ for a pure and
infinite \oneD\ system. As $\eta \to 0$, Eq.(\ref{eq:pureDoS}) displays
the discrete behavior for the \dos\ of a clean ring.

\begin{figure}
  \centerline{\epsfxsize10cm \epsfbox{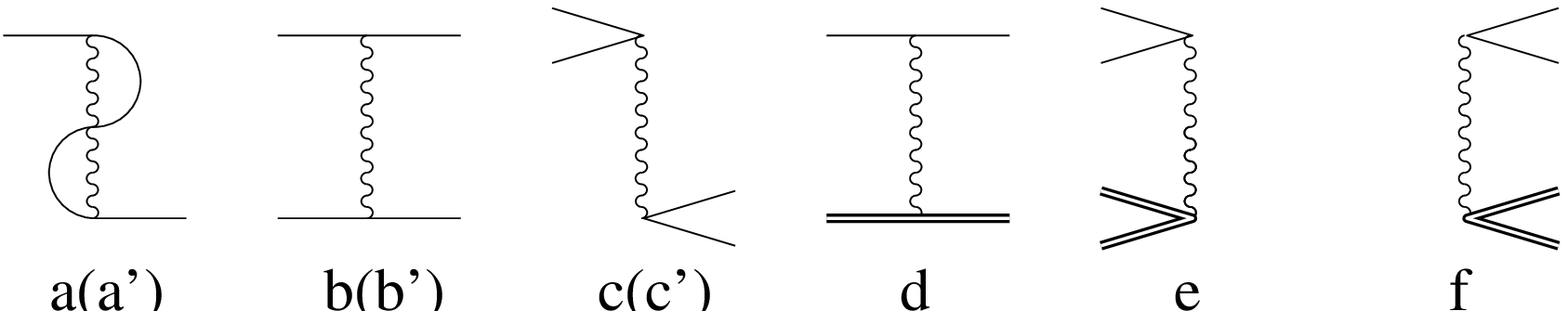}}
\vspace{0.2cm}
\caption{Contributing internal vertices. Vertices a',b', and c' have the
  same form as a,b, and c, but double lines instead of single ones.
  The following factors correspond to the vertices: a,a' $\sim$
  $(\frac{-1}{2 \lm} - \frac{1}{2 \lp})$; b,b' $\sim $ $\frac{-1}{\lp}$;
  c,c' $\sim$ $\frac{-1}{\lm}$; d $\sim$ $\frac{+1}{\lp}$; e $\sim$
  $\frac{+1}{\lm}\exp(\frac{4\eta x}{v(\epsilon)})$; f $\sim$
  $\frac{+1}{\lm}\exp(\frac{-4\eta x}{v(\epsilon)})$}
\label{fig:vertices}
\end{figure}

\begin{figure}
\begin{align*}
  \epsfbox{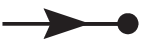} &\;\sim\; \sqrt{-\frac{i}{v}} e^{+\frac{2 \pi i}{L}
    \frac{\phi}{\phi_{0}} x - i p x + \frac{\eta}{v} x} &
  \epsfbox{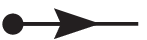} &\;\sim\; \sqrt{-\frac{i}{v}} e^{-\frac{2 \pi i}{L}
    \frac{\phi}{\phi_{0}} x + i p x
    - \frac{\eta}{v} x} \\
  \epsfbox{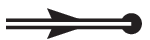}& \;\sim\; \sqrt{\frac{i}{v}} e^{+\frac{2 \pi i}{L}
    \frac{\phi}{\phi_{0}} x + i p x + \frac{\eta}{v} x} &
  \epsfbox{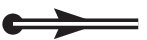} &\;\sim\; \sqrt{\frac{i}{v}} e^{-\frac{2 \pi i}{L}
    \frac{\phi}{\phi_{0}} x - i p x
    - \frac{\eta}{v} x} \\
  \epsfbox{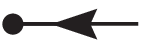}&\;\sim\; \sqrt{-\frac{i}{v}} e^{+\frac{2 \pi i}{L}
    \frac{\phi}{\phi_{0}} x + i p x - \frac{\eta}{v} x} &
  \epsfbox{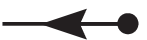} &\;\sim\; \sqrt{-\frac{i}{v}} e^{-\frac{2 \pi i}{L}
    \frac{\phi}{\phi_{0}} x - i p x
    + \frac{\eta}{v} x} \\
  \epsfbox{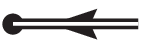}&\;\sim\; \sqrt{\frac{i}{v}} e^{+\frac{2 \pi i}{L}
    \frac{\phi}{\phi_{0}} x - i p x - \frac{\eta}{v} x} &
  \epsfbox{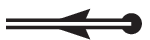} &\;\sim\; \sqrt{\frac{i}{v}} e^{-\frac{2 \pi i}{L}
    \frac{\phi}{\phi_{0}} x + i p x
    + \frac{\eta}{v} x} \\
\end{align*}
\caption{The external vertices. One incoming vertex and one outgoing vertex are attached to each continuous fermion line characterizing one Green's function.}
\label{fig:ext_vertices}
\end{figure}

In the diagrammatical technique, the retarded (advanced) GF of an
electron moving in the field of randomly distributed impurities is
represented by an ordinary (double) continuous line in real space,
which goes from point $x$ to $x'$ after multiple scattering on a given
impurity configuration, realized by the potentials
$V(x_i)$ with impurities placed at the points
$\{x_i\}$. For the \dos\ and its moments it suffices to consider $x=x'$.
Since the ``bare'' GFs [Eq.(\ref{eq:bareGF})] between two successive
scatterings have factorable structure, the coordinate dependence can
be transfered from the lines to the vertices. Averaging over the
random Gaussian potential leads to a pairing of the impurity vertices.
Their strength is measured by the inverse forward (backward)
scattering length,
\begin{equation}
\frac{1}{\lp(\epsilon)}=\frac{2}{\hbar^2 v^2(\epsilon)}\int_0^\infty U(x)dx \quad
\mbox{and}\quad \frac{1}{\lm(\epsilon)}=\frac{2}{\hbar^2 v^2(\epsilon)}\int_0^\infty
U(x) dx \cos (p(\epsilon)x)
\end{equation}
For the Born approximation to be applicable, the correlator $U(x-x')=\avg{V(x)V(x')}$
should have a width much smaller than the mean distance between
impurities $\frac{1}{c}$, and $\frac{1}{c} \ll \l^\pm$.
In the extreme case of a white noise
potential $U(x-x')=c U_0^2 \sum_{n=-\infty}^{\infty} \delta(x-x'+n
L)$, the two scattering lengths become equal, $\lm=\lp=2 l$.  Given in
Fig.\ref{fig:vertices} are the essential vertices selected according
to the condition $p_Fl \gg 1$, with $p_F$ and $l$ being the Fermi momentum and
the mean free path, respectively. Although the ``bare'' GFs depend on
the direction due to the magnetic field, the internal vertices in
Fig.\ref{fig:vertices} do not differ from those of Berezinskii. All
dependence on the magnetic field is transfered from the lines to the
external vertices, which are shown in Fig.\ref{fig:ext_vertices}.

\begin{figure}
  \centerline{\epsfxsize14cm \epsfbox{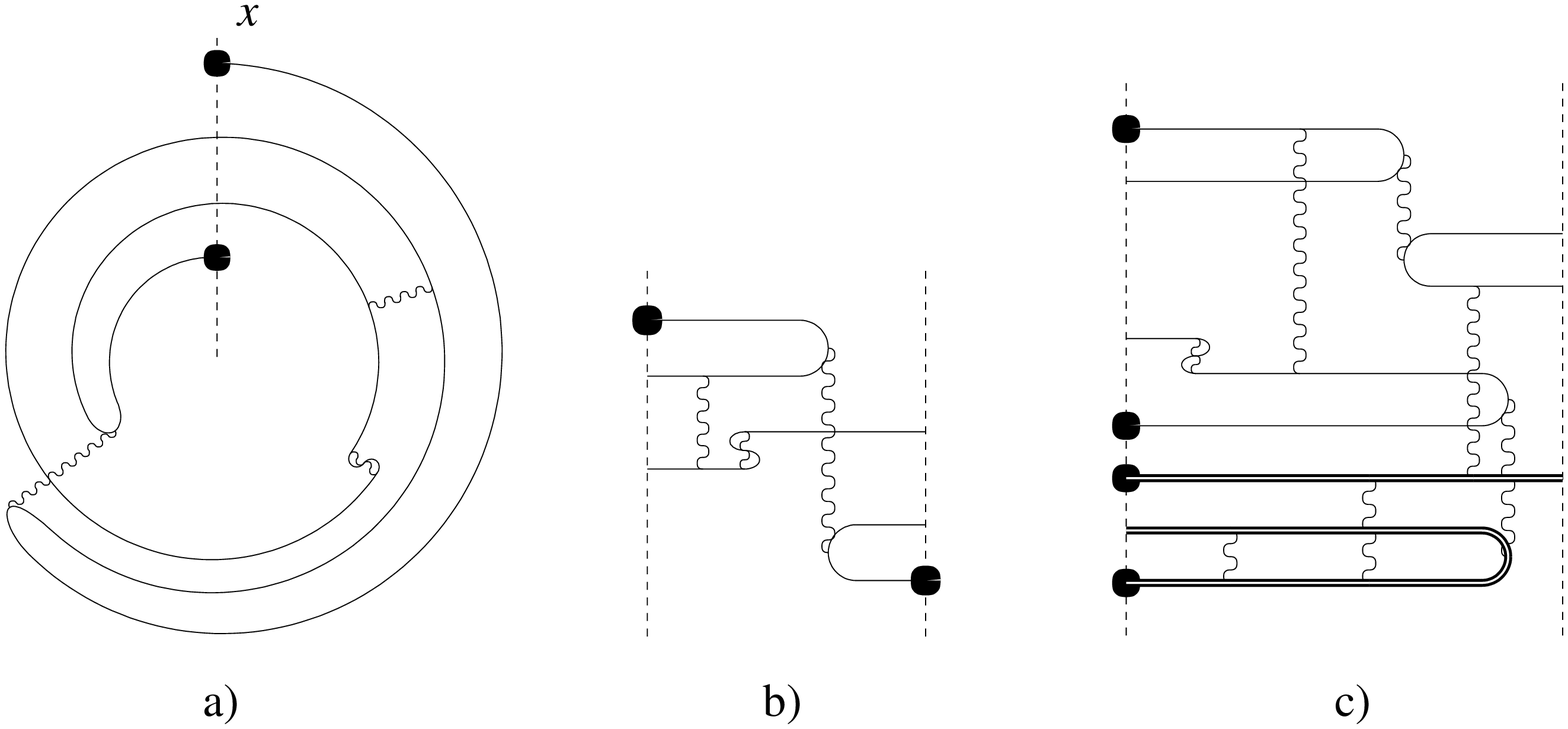}}
\vspace{0.3cm}
\caption{A simple diagram contributing to the single GF $\avg{G_R}$ a) resembling the motion of the electron around the ring and b) after cutting at point $x$. The numbers characterizing this diagram are $m=1$ and $n=1$. c) shows a contribution to $\avg{G_RG_A}$. Here, $m=2$, $m'=1$, $n=0$, $\mb=1$, $\mb'=0$, and $\nb=1$.}
\label{fig:diagramExample}
\end{figure}

 As an example, a simple diagram contributing to $\avg{G_R(\epsilon,\phi;x,x)}$
is drawn in Fig.\ref{fig:diagramExample}a. 
For convenience, we cut the
diagram at point $x$ and straighten the lines, which results in
Fig.\ref{fig:diagramExample}b. Each diagram for a single GF 
is then characterized by $m$ pairs of lines returning to $x$ and $n$ through-going lines. Since the bulk of
each diagram (i.e.\ after the removal of the external vertices) does
not depend on the direction, $n$ is the sum of right-going
$\np$ and left-going $\nm$ lines.
For correlators $\avg{G_R^l(\epsilon,\phi;x,x)G_A^{k-l}(\epsilon,\phi;x,x)}$,
we have to distinguish the number $m$ of returning lines on the l.h.s.
and the number $m'$ of returning lines on the r.h.s., and in addition to 
introduce $\mb$, $\mb'$, and $\nb$ for the advanced GF. An example for this case is shown in Fig.\ref{fig:diagramExample}c. 

In contrast to Berezinskii's technique for an infinite \oneD\ system,
the external vertices depend on the direction. Also, the
diagrams carry a factor $\exp[2 \pi i \frac{\phi}{\phi_0} (\np+\npb-\nm-\nmb)]$ from
the through-going lines. 
The number of pairs on the two sides of the cutting line, which we denote
by $m$ and $m'$ for the retarded GF and by $\mb$ and $\mb'$ for the advanced
GF, may in general differ by $\pm 1$ (i.e.\ $m-m'=-1,0,1$ and $\mb-\mb'=-1,0,1$) as in Sec.~\ref{sec:moments}. However, for $\avg{G_R(\epsilon,\phi;x,x')}$ in the regime of weak disorder as in Sec.~\ref{sec:dos}, we have always $m=m'$.

\begin{figure}
  \centerline{\epsfxsize12cm \epsfbox{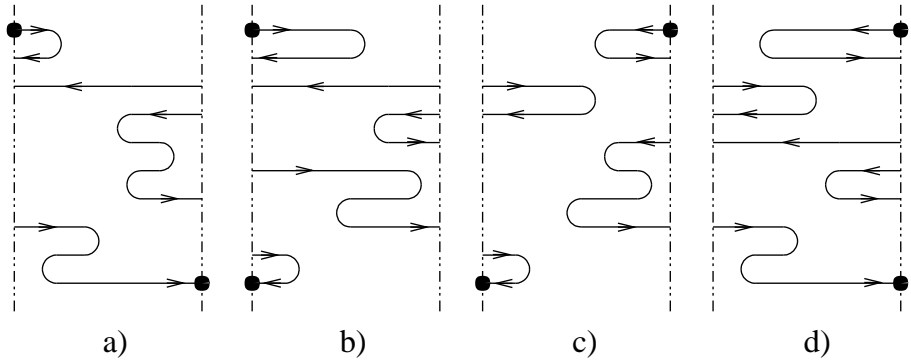}}
\vspace{0.3cm}
\caption{The different possibilities to attach external vertices. The
  cases b) and d) contain oscillating factors $\exp(\pm 2 i
  p(\epsilon)x)$, which is connected to the fact that
  the numbers of line pairs on the left and
  right hand side differ by $\pm 1$.  The cases a) and c) have $m=m'$
  and hence do not contain such factors.}
\label{fig:attach_ext_vertices}
\end{figure}

\section{The density of states}
\label{sec:dos}
The diagrams for the \dos\ do not exhibit the full complexity presented
in the previous section, since they contain only retarded GFs.
Consequently, we can omit vertices d),e), and f) of
Fig.\ref{fig:vertices} and set $m=m'$. Following Berezinskii's method
,\cite{berezinskii73a} we denote the sum of all diagrams with $m$
pairs of returning lines and $n=\np+\nm$ through-going lines by
$Q_{0}(m,n;x-x'=L)$, where $\np$($\nm$) is the number of
right-(left-)going lines that cross the whole diagram. 
Such a diagram is shown in Fig.\ref{fig:diagramExample}a and \ref{fig:diagramExample}b. Since the
magnetic field dependence has been extracted from $Q_{0}$, it only depends
on the total number of through-going lines.  The condition $m=m'$
restricts the possibilities to attach the external vertices to the
cases a) and c) of Fig.\ref{fig:attach_ext_vertices}.  The average
value of the retarded GF can be expressed in terms of the kernel
$Q_{0}$ as
\begin{equation}
        \begin{split}
          \avg{G^+(\epsilon,\phi;x,x)}=&-\frac{i}{v(\epsilon)}
          \sum_{m=0}^\infty \sum_{\np=0}^\infty \sum_{\nm=0}^\infty
          \Bigl[
          \binom{m+\np}{m}\binom{m-1+\nm}{m-1}+\binom{m+\nm}{m}\binom{m-1+\np}{m-1}
          -\delta_{m,0}\delta_{\np,0}\delta_{\nm,0}\Bigr]\\
          &\exp\bigl\{i p(\epsilon) L(\np+\nm)-\frac{\eta
            L}{v(\epsilon)}(\np+\nm)-2 \pi i
          \frac{\phi}{\phi_{0}}(\np-\nm)\bigr\} Q_{0}(m,\np+\nm;L)
\end{split}
\label{eq:singleGF}
\end{equation}
The two products of binomials in the bracket of Eq.(\ref{eq:singleGF})
characterize the different possibilities to insert the $\np$
right-going and the $\nm$ left-going lines between $m$ pairs and
correspond to the cases a) and c) of
Fig.\ref{fig:attach_ext_vertices}, respectively. For $m=\np=\nm=0$,
these two possibilities are degenerated into a point--like diagram.
To avoid double--counting in this case, the third term in the brackets
has been added. The combinatorial factor, corresponding e.g.\ to the
configuration in Fig.\ref{fig:attach_ext_vertices}a, can be obtained
as follows: $\np$ lines can be distributed at $m+1$ positions, before
each of the loops on the l.h.s.\ and after the last loop, whereas the
$\nm$ left--going lines can be inserted at $m$ positions before each of
the loops on the r.h.s.. Denoting the number of lines at a given position
with $n_i^\pm$, we have the restrictions $\np = \np_1+\np_2\dots+\np_{m+1}$
and $\nm = \nm_1+\nm_2\dots +\nm_m$.

Summing over all these possibilities gives
\begin{equation}
        \sum_{\{\np_{i}\}}\delta_{\np,\np_{1}+\np_{2}+\dots+\np_{m+1}}
        \sum_{\{\nm_{i}\}}\delta_{\nm,\nm_{1}+\nm_{2}+\dots+\nm_{m}}
        = \binom{m+\np}{m}\binom{m-1+\nm}{m-1}
\label{eq:binomials}
\end{equation}
The expression for the \dos\ can be obtained by combining
Eqs.(\ref{eq:dos_formula}) and (\ref{eq:singleGF}):
\begin{equation}
        \begin{split}
          \rho(\epsilon,\phi)=\rho_{0}\sum_{m=0}^{\infty}\sum_{n=0}^
{\infty}\sum_{k=0}^{n}
          \Bigl[2\binom{m+k}{m}\binom{m+n-k-1}{m-1}-\delta_{m,0}
\delta_{n,0}\Bigr]\\
          \cos(p(\epsilon)Ln)\exp(-\frac{\eta}{v(\epsilon)}Ln)\cos(2
          \pi \frac{\phi}{\phi_{0}}(2k-n))Q_{0}(m,n;L)
        \end{split}
\label{eq:rho}
\end{equation}

\begin{figure}
  \centerline{\hspace{1cm} \epsfxsize12cm \epsfbox{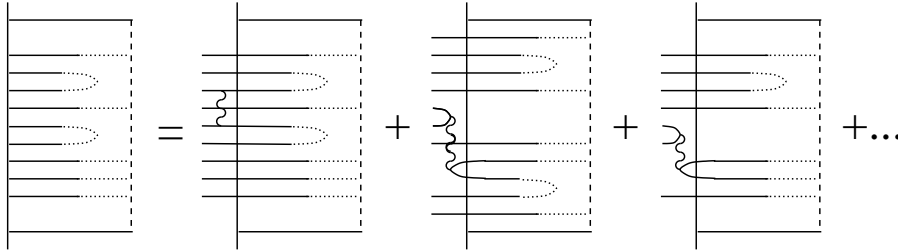}}
\caption{How to include the internal vertices to $Q_0$. Shown on the r.h.s. 
is the inclusion of vertex b) of Fig.\ref{fig:vertices} and the inclusion of vertex c) without and with changing of $m$ and $n$.}
\label{fig:Q0_inclusion}
\end{figure}

To find $Q_{0}(m,n;x-x')$, we shift $x$ infinitesimally and examine
the different impurity vertices that pass through the point $x$. The result is a differential equation for $Q_0$,
\begin{equation}
        \frac{\dif}{\dif x} 
        Q_{0}(m,n;x)=-\bigl[\frac{(2m+n)^2}{2\lp}+\frac{n}{2 
        \lm}+\frac{m(m+n)}{\lm}\bigr]Q_{0}(m,n;x)-\frac{1}{\lm}m^2Q_0(m-1,n+2;x)
        \label{eq:DGL_for_Q0}
\end{equation}
where the vertices a),b), and c) in Fig.\ref{fig:vertices} contribute.
For a) the number of possibilities to be included is $2m+n$, for b)
$\frac12 (2m+n)(2m+n-1)$ and to include c) there are $m(m+n-1)$ ways
that do not change $m$ and $n$ and $m^2$ possibilities that decrease
$m$ by 1 pair and increase $n$ by 2. The process of construction of this equation  is illustrated in Fig.\ref{fig:Q0_inclusion}.

The kernel $Q_0$ satisfies the boundary condition
\begin{equation}
        Q_0(m,n;x-x'=L=0)=\delta_{m,0}
        \label{eq:BC_for_Q0}
\end{equation}
which expresses the absence of scattering for a ring with
infinitesimal small circumference.  The Eq.(\ref{eq:DGL_for_Q0}) for
$Q_0(m,n;x)$ can be solved exactly as shown in
Appendix~\ref{app:Calc_of_Q0}:
\begin{equation}
\begin{split}
Q_0(m,n;L)= &\exp\bigl\{- \frac{L}{2\lp}(2m+n)^2-\frac{L}{\lm}m(m+n)-
\frac{L}{2\lm}n\bigr\}\\
&\sum_{j=0}^{m}(-1)^j\binom{m}{j} \frac{m!(j+n-2)!}{(m+j+n-1)!}(2j+n-1)
          \exp\bigl\{ \frac{L}{\lm} j(j+n-1)\bigr\}
\end{split}
\label{eq:Q0_sol}
\end{equation}
Equations (\ref{eq:rho}) and (\ref{eq:Q0_sol}) give a complete
description of the \dos\ for a one--channel ring in an external magnetic
field. They are exact in the regime of weak disorder, when the
condition $p_F l^\pm \gg 1$ or $\epsilon_F \tau \gg 1$ is satisfied
($p_F$ and $\epsilon_F$ are Fermi momentum and Fermi energy,
respectively). Further, since the circumference $L$ of the ring may
vary within a large range ($L \leq l^\pm$ to $l^\pm \ll L < \infty$),
the results cover both the weak localization and the ballistic
regimes.  For the weak localization regime, when $L \gg \max(\lp,\lm)$
the amplitude of $Q_0(m,n;L)$ decreases rapidly with $m$ and $n$ due
to the exponential prefactor. Keeping harmonics up to $n=2$ in
Eq.(\ref{eq:rho}) and using Eq.(\ref{eq:Q0star_n0}), we obtain the
leading behavior for $\rho(\epsilon,\phi)$
\begin{equation}
\begin{split}
  \rho(\epsilon,\phi)= &
\rho_0\bigl[1-\frac{2L}{\lm}\exp(-\frac{2L}{\lp}-\frac{L}{\lm})\bigr]
  +2\rho_0 \exp(-\frac{L}{2\lp}-
\frac{L}{2\lm})\Bigl[\cos(p(\epsilon)L)
\cos(2 \pi \frac{\phi}{\phi_0})e^{-\frac{\eta L}{v(\epsilon)}}\\
&  +e^{-\frac{3L}{2\lp}-\frac{L}{2\lm}}\cos(2p(\epsilon)L)\cos(4 \pi
  \frac{\phi}{\phi_0})e^{-\frac{2\eta}{v(\epsilon)}L}\Bigr]
\end{split}
\label{eq:rho_weak_localization}
\end{equation}
We see that the main quantum correction to the average value of the
\dos\ oscillates with a period of $\phi_0$. The amplitude of this
contribution decreases exponentially with the impurity strength or
with increasing $L$, so that $\rho=\rho_0$ for $L \to \infty$.

The ballistic regime is realized for $L \leq \min(\lp,\lm)$. Keeping
terms up to first order in $m$, the \dos\ can be approximated in this
limit by
\begin{equation}
\begin{split}
  \rho(\epsilon,\phi)=\rho_0(\epsilon,\phi)-
\rho_0\frac{L}{\lp}
\sum_{n=0}^{N^+}n^2\cos(p(\epsilon)Ln)\cos(2 \pi \frac{\phi}{\phi_0}n) - 
\rho_0\frac{L}{\lm}\sum_{n=0}^{N^-}n\cos(p(\epsilon)Ln) 
\cos(2\pi \frac{\phi}{\phi_0}n)\\
  - 2 \rho_0 \frac{L}{\lm}\sum_{n=0}^{N^-}\sum_{k=0}^n (k+1)
  \cos(p(\epsilon)Ln)\cos(2 \pi \frac{\phi}{\phi_0}(2k-n))
\end{split}
\end{equation}
where $N^+\approx\bigl[\sqrt{\frac{2\lp}{L}}\bigr]$ and
$N^-\approx\bigl[\frac{2\lm}{L}\bigr]$, and the \dos\ of a clean ring
$\rho_0(\epsilon,\phi)$ is given by Eq.(\ref{eq:pureDoS}). In the
ballistic regime, $L$ is of the same order of magnitude as $l^\pm$,
hence $N^\pm$ may be rather small integers. Therefore the oscillation
with the full flux quantum $\phi_0$ dominates in the ballistic regime.

In the absence of backward scattering ($\lm=\infty$), $Q_0(m,n;L)$ is
greatly simplified:
\begin{equation}
Q_0(m,n;L)=\exp\bigl\{-\frac{L}{2\lp}n^2\bigr\}\delta_{m,0}
\label{eq:Q0_no_backward_scattering}
\end{equation}
Substituting Eq.(\ref{eq:Q0_no_backward_scattering}) in (\ref{eq:rho})
and using $\lim_{m \to 0}\binom{m+n-k-1}{m-1}=\delta_{n,k}$, we can
express the \dos\ as

\begin{equation}
\begin{split}
  \rho(\epsilon,\phi)=\rho_0+\frac{\rho_0}{2}\sqrt{\frac{\lp}{2 \pi L}}
\int_{-\infty}^{\infty}
  \dif z
  e^{-\frac{\lp}{2L}z^2}\Bigl\{
\frac{1}{\exp(-ip(\epsilon)L-i2\pi\phi/\phi_0+iz+
    \frac{\eta}{v(\epsilon)}L)-1} \\
  +
  \frac{1}{\exp(-ip(\epsilon)L+i2\pi\phi/\phi_0+iz
+\frac{\eta}{v(\epsilon)}L)-1}
  +c.c \Bigr\}
\end{split}
\label{eq:rho_no_backward_scattering}
\end{equation}
From Eq.(\ref{eq:rho_no_backward_scattering}) one sees that forward
scattering coherently shifts all energy levels. The value of this
shifting is random with Gaussian distribution, with a typical value of
$\frac{\hbar}{\tau^+}\sqrt{\frac{\lp}{L}}$, where $\tau^+$ is the
relaxation time due to forward scattering. The level repulsion\cite{altshuler86a}
in \oneD\ disordered systems therefore is only due to backward
scattering. The width of this level broadening of
the averaged system is much smaller than Dingle broadening for the
weak localization regime, whereas the two mechanisms can have
comparable effects in the ballistic regime.

\section{Higher moments of the \dos\ and distribution functions}
\label{sec:moments}

Exact calculations show that the average value of the \dos\ oscillates
with a period of the flux quantum $\phi_0$. To understand the reason
for the experimentally observed oscillation of the persistent current
in a sufficiently large ring ($L\gg l$)\cite{levy90a,mailly95a} with the halved period,
we calculate here higher moments of the distribution of the \dos.
According to Eq.(\ref{eq:rhok_definition}), we have to determine the
correlators $\avg{G_R^l G_A^{k-l}}$ for the $k$th moment. In contrast
to the Berezinskii technique for strictly \oneD\ systems, the correlators
here are characterized by only one block $Q$. An example is shown in
Fig.\ref{fig:diagramExample}c.  Each diagram contributing to
$\avg{G_R^l G_A^{k-l}}$ consists of $l$ retarded and $k-l$ advanced
lines. For the $i$th retarded (advanced) GF, we count the number of left loops
$m_i$($\mb_i$), of right loops $m'_i$($\mb'_i$), and of left- and
right-going traversing lines $n_i=\np_i+\nm_i$($\nb_i=\npb_i+\nmb_i$).
(Notice that the index $i$ of $n_i$ here 
denotes the number of the GF, whereas it
was used for the position within one GF in the previous section.)
For the different fermion lines, we now can
attach the external vertices in four different ways, as shown in
Fig.\ref{fig:attach_ext_vertices}. This allows $m_i-m'_i=-1,0,1$ 
($\mb_i-\mb'_i=-1,0,1$) for the each retarded (advanced) GF.  It
turns out that the block $Q$ depends only on the total numbers
$m=m_1+m_2+\dots+m_l$, $\mb=\mb_1+\mb_2+\dots+\mb_{k-l}$ and similarly
$m'$,$\mb'$,$n$,$\nb$[\onlinecite{altshuler89a}].  It is clear from these
considerations that the difference between $m$ and
$m'$ is always restricted by $-l\leq m-m' \leq l$. This is
automatically taken into account by the mixing coefficient\cite{altshuler89a}
$\varphi_l(m,m';\np,\nm)$. This coefficient is the generalization of
the term in brackets of Eq.(\ref{eq:singleGF}). It counts the
different possibilities to attach external vertices (see
Fig.\ref{fig:attach_ext_vertices}) and to distribute the through-going
lines between the loops. Using Eq.(\ref{eq:binomials}), we can write
it as
\begin{equation}
\begin{split}
  \varphi_l(m,m';&\np,\nm)= \sum_{\{m_i\}}\delta_{m,m_1+m_2+\dots+m_l}
  \sum_{\{m'_i\}}\delta_{m',m'_1+m'_2+\dots+m'_l}
  \sum_{\{\np_i\}}\delta_{\np,\np_1+\np_2+\dots+\np_l}
  \sum_{\{\nm_i\}}\delta_{\nm,\nm_1+\nm_2+\dots+\nm_l}\\
&  \prod_{i=1}^l\Bigl\{ e^{2 i p(\epsilon)x-\frac{2
      \eta}{v(\epsilon)}x}
  \delta_{m_i,m'_i+1}\binom{m_i-1+\np_i}{m_i-1}\binom{m_i-1+\nm_i}{m_i-1}
  +e^{-2 i p(\epsilon)x+\frac{2 \eta}{v(\epsilon)}x}
  \delta_{m_i,m'_i-1}\binom{m_i+\np_i}{m_i}\binom{m_i+\nm_i}{m_i}\\
&  +\delta_{m_i,m'_i}\bigl[\binom{m_i+\np_i}{m_i}\binom{m_i-1+\nm_i}{m_i-1}+
  \binom{m_i-1+\np_i}{m_i-1}\binom{m_i+\nm_i}{m_i}
  -\delta_{m_i,0}\delta_{\np_i,0}\delta_{\nm_i,0}\bigr]\Bigr\}
\end{split}
\label{eq:comb}
\end{equation}
Separating the exponential factors in Eq.(\ref{eq:comb}), we can write
$\varphi_l$ as
\begin{equation}
\varphi_l(m,m';\np,\nm)=e^{-2 i x(m'-m)(p(\epsilon)+i\frac{\eta}{v(\epsilon)})}
\phit_l(m,m';\np,\nm)
\label{eq:mixing_substitution}
\end{equation}
with
\begin{equation}
\begin{split}
  \phit_l(m,m';\np,\nm)= & \sum_{\{m_i\}}\delta_{m,m_1+m_2+\dots+m_l}
  \sum_{\{m'_i\}}\delta_{m',m'_1+m'_2+\dots+m'_l}
  \sum_{\{\np_i\}}\delta_{\np,\np_1+\np_2+\dots+\np_l}
  \sum_{\{\nm_i\}}\delta_{\nm,\nm_1+\nm_2+\dots+\nm_l}\\
&  \prod_{i=1}^l\Bigl\{
  \delta_{m_i,m'_i+1}\binom{m_i-1+\np_i}{m_i-1}\binom{m_i-1+\nm_i}{m_i-1}
  + \delta_{m_i,m'_i-1}\binom{m_i+\np_i}{m_i}\binom{m_i+\nm_i}{m_i}\\
&  +\delta_{m_i,m'_i}\bigl[\binom{m_i+\np_i}{m_i}\binom{m_i-1+\nm_i}{m_i-1}+
  \binom{m_i-1+\np_i}{m_i-1}\binom{m_i+\nm_i}{m_i}
  -\delta_{m_i,0}\delta_{\np_i,0}\delta_{\nm_i,0}\bigr]\Bigr\}
\end{split}
\label{eq:comb_transformed}
\end{equation}
The mixing factor for the advanced GF is obtained from complex
conjugation of $\varphi_{k-l}$.

Now we can express Eq.(\ref{eq:rhok_definition}) in terms of $Q$ and
the mixing factors
\begin{equation}
\begin{split}
  \avg{\rho^k(\epsilon,\phi;L,x)} =& \bigl(\frac{\rho_0}{2}\bigr)^k
  \sum_{l=0}^k \sum_{m=0}^{\infty}\sum_{\mb=0}^{\infty}
  \sum_{m'=0}^{\infty}\sum_{\mb'=0}^{\infty}
  \sum_{\np=0}^{\infty}\sum_{\nm=0}^{\infty}
  \sum_{\npb=0}^{\infty}\sum_{\nmb=0}^{\infty}
  \binom{k}{l}\\
  &e^{-i 2 \pi \frac{\phi}{\phi_0}(\np-\nm+\npb-\nmb)}
  e^{i p(\epsilon)L (\np+\nm-\npb-\nmb)}
  e^{- \frac{\eta L}{v(\epsilon)} (\np+\nm+\npb+\nmb)}\\
  &e^{-2ip(\epsilon)x(m'-m-\mb'+\mb)}
  e^{\frac{2\eta x}{v(\epsilon)}(m'-m+\mb'-\mb)}\\
  &\phit_l(m,m';\np,\nm) \phit_{k-l}(\mb,\mb';\npb,\nmb) Q\Bigl(
\begin{matrix}
  m,&m',&\np+\nm\\
  \mb,&\mb',&\npb+\nmb
\end{matrix}
\Bigr|L\Bigr)
\end{split}
\label{eq:rhok}
\end{equation}
Here, the first three exponential factors come from the external vertices and
from the revolutions around the ring. The last two exponential factors
were separated from the mixing coefficients [Eq.(\ref{eq:mixing_substitution})].
Below, we shall see that the last exponential factor is canceled by a contribution from $Q$.

\subsection{Equation for the central block $Q$}

As noted above, the central block $Q\Bigl(
\begin{matrix}
  m,&m',&n\\
  \mb,&\mb',&\nb
\end{matrix}
\Bigr|x\Bigr)$ is defined as the sum of all diagrams with $m$ and
$\mb$ pairs of returning lines on the left side, $m'$ and $\mb'$ pairs
on the right side and $n$ and $\nb$ through-going lines, coming from
retarded and advanced GFs, respectively. An equation determining $Q$
can be constructed according to Berezinskii's idea by attaching all
possible vertices, given in Fig.\ref{fig:vertices}, to the existing
block, while avoiding the formation of unconnected electron loops.
Careful analysis of all these possibilities gives the equation
\begin{equation}
\begin{split}
  \frac{\dif}{\dif x} & Q\Bigl(
\begin{matrix}
  m,&m',&n\\
  \mb,&\mb',&\nb
\end{matrix}
\Bigr|x\Bigr) =-\Bigl[\frac{1}{2\lp}(2m+n-2\mb-\nb)^2 +
\frac{1}{\lm}m(m+n) + \frac{1}{\lm}\mb(\mb+\nb) +
\frac{1}{2\lm}(n+\nb)\Bigr]
Q\Bigl(\begin{matrix} m,&m',&n\\ \mb,&\mb',&\nb \end{matrix} \Bigr|x\Bigr)\\
&-\frac{1}{\lm}mm' Q\Bigl(\begin{matrix} m-1,&m'-1,&n+2\\ 
  \mb,&\mb',&\nb \end{matrix} \Bigr|x\Bigr) -\frac{1}{\lm}\mb\, \mb'
Q\Bigl(\begin{matrix} m,&m',&n\\ \mb-1,&\mb'-1,&\nb+2 \end{matrix} \Bigr|x\Bigr)\\
&+\frac{1}{\lm}m\mb e^{\frac{4\eta x}{v(\epsilon)}}
Q\Bigl(\begin{matrix} m-1,&m',&n\\ \mb-1,&\mb',&\nb \end{matrix} \Bigr|x\Bigr)\\
&+\frac{1}{\lm}(m+n)(\mb+\nb) e^{-\frac{4\eta x}{v(\epsilon)}}
Q\Bigl(\begin{matrix} m+1,&m',&n\\ \mb+1,&\mb',&\nb \end{matrix}
\Bigr|x\Bigr) +\frac{1}{\lm}m'(\mb+\nb) e^{-\frac{4\eta
    x}{v(\epsilon)}}
Q\Bigl(\begin{matrix} m,&m'-1,&n+2\\ \mb+1,&\mb',&\nb \end{matrix} \Bigr|x\Bigr)\\
&+\frac{1}{\lm}(m+n)\mb' e^{-\frac{4\eta x}{v(\epsilon)}}
Q\Bigl(\begin{matrix} m+1,&m',&n\\ \mb,&\mb'-1,&\nb+2 \end{matrix}
\Bigr|x\Bigr) +\frac{1}{\lm}m'\mb' e^{-\frac{4\eta x}{v(\epsilon)}}
Q\Bigl(\begin{matrix} m,&m'-1,&n+2\\ \mb,&\mb'-1,&\nb+2 \end{matrix}
\Bigr|x\Bigr)
\end{split}
\label{eq:Q_DGL}
\end{equation}

\begin{figure}
  \centerline{\epsfxsize18cm \epsfbox{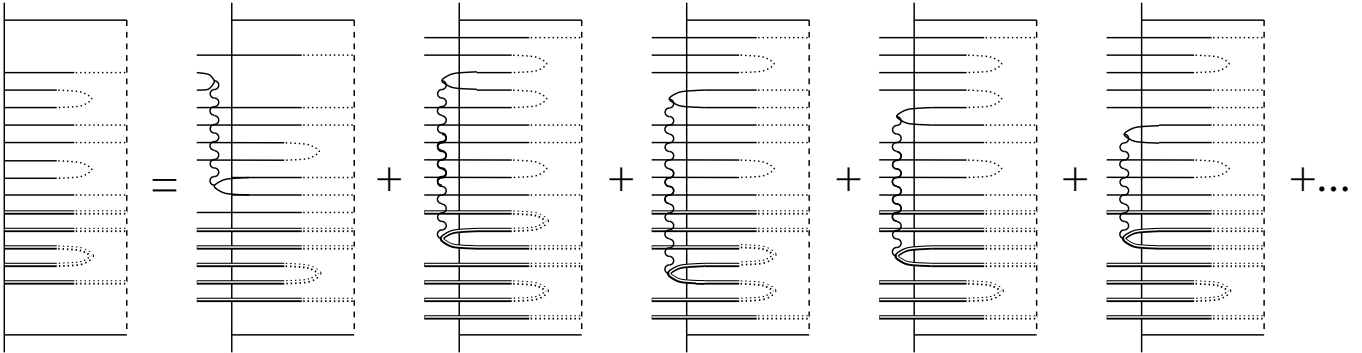}}
\caption{Different possibilities to attach vertices to the block $Q$,
  as the left side is shifted. The first plot shows the inclusion of
  vertex c) of Fig.\ref{fig:vertices} with decreasing $m$ and $m'$ by
  1 and increasing $n$ by 2.  The other four blocks show the different
  possibilities to include vertex f) and correspond to the last four
  terms in Eq.(\ref{eq:Q_DGL}).}
\label{fig:include_f}
\end{figure}

The block $Q$ is subjected to a similar boundary condition as $Q_0$ in
the previous section
\begin{equation}
Q\Bigl(\begin{matrix} m,&m',&n\\ \mb,&\mb',&\nb \end{matrix} \Bigr|x=0\Bigr)
=\delta_{m,0}\delta_{m',0}\delta_{\mb,0}\delta_{\mb',0}
\label{eq:Q_BC}
\end{equation}
which states that for an infinitesimal ring there can be no
scattering.  The first coefficient on the right hand side of
Eq.(\ref{eq:Q_DGL}) contains contributions from the vertices
a),a'),b),b'),c),c'), and d) in Fig.\ref{fig:vertices}. Vertices
a),b), and d) can be attached in $(2m+n)$,$\frac12(2m+n)(2m+n-1)$, and
$(2m+n)(2\mb+\nb)$ ways, respectively, the coefficients for a') and
b') are the same as for a) and b), with the replacement $\{m,n\} \to
\{\mb,\nb\}$.  For vertex c) we have again to distinguish two
possibilities as in Sec.\ref{sec:dos}. We have $m(m+n-1)$ ways to
attach it without changing $m$,$m'$, and $n$; and $mm'$ different ways
with changing $\{m,m'n\} \to \{m-1,m'-1,n+2\}$.  The latter kind of
insertion of the vertex c) in Fig.\ref{fig:vertices}
and its counterpart for the advanced GF give the
second and the third term on the right hand side of
Eq.(\ref{eq:Q_DGL}).  The inclusion of vertex e) reduces $m$ and $\mb$
by 1.  The insertion of vertex f), however, can be done in four
different ways which are shown schematically in the last four blocks of
Fig.\ref{fig:include_f}.

Trying to solve Eq.(\ref{eq:Q_DGL}), one may begin by substituting
$\tilde{Q}=\frac{1}{m'!\mb'!}Q$ and then introduce new variables
$M=2m+n$, $\overline{M}=2\mb+\nb$, $M'=2m'+n$, $\overline{M}'=2\mb'+\nb$.
As a consequence, $M'$ and $\overline{M}'$ appear as fixed parameters
in the differential equation for $\tilde{Q}$. But still then,
$\tilde{Q}$ depends on the five variables
$M$,$n$,$\overline{M}$,$\nb$, and $x$.  Under these circumstances,
looking for the general analytic solution of Eq.(\ref{eq:Q_DGL}), one
meets with enormous difficulties. Before studying an asymptotic
approximation of the problem, we make some simplifications of
Eqs.(\ref{eq:rhok}) and (\ref{eq:Q_DGL}).  The exponential factors in
Eq.(\ref{eq:Q_DGL}) can be removed by the following substitution:
\begin{equation}
Q\Bigl(\begin{matrix} m,&m',&n\\ \mb,&\mb',&\nb \end{matrix} \Bigr|x\Bigr)=\exp\bigl\{\frac{2\eta x}{v(\epsilon)}(m+\mb-m'-\mb')\bigr\}
\overline{Q}\Bigl(\begin{matrix} m,&m',&n\\ \mb,&\mb',&\nb \end{matrix} \Bigr|x\Bigr)
\label{eq:Q_substitution}
\end{equation}

The equation for $\overline{Q}$ has the same structure as Eq.(\ref{eq:Q_DGL}), 
only the exponential factors are dropped and the first term on the r.h.s.\ of
Eq.(\ref{eq:Q_DGL}) acquires another contribution $-\frac{2\eta}{v(\epsilon)}(m+\mb-m'-\mb')$ to the prefactor of $Q$.

From the structure of the internal vertices in Fig.\ref{fig:vertices} one sees
that the condition 
\begin{equation}
(m'-m)-(\mb'-\mb)=0
\label{eq:m-m'-connection}
\end{equation}
is satisfied for arbitrary cross sections. (The same condition also applies to 
the strictly \oneD\ problem, see [\onlinecite{altshuler89a}].) A corresponding symmetry of Eq.(\ref{eq:Q_DGL}) confirms this condition.

In the regime of weak disorder, $p_F l \gg 1$, Eq.(\ref{eq:rhok}) for
the $k$th moment of the \dos\ contains terms that strongly oscillate
with the particle energy ($\np + \nm \neq \npb + \nmb$) apart
from smooth ones ($\np + \nm = \npb + \nmb$). To neglect the strongly oscillating terms, we choose only those terms in Eq.(\ref{eq:rhok}) that satisfy the condition

\begin{equation}
\np + \nm = \npb + \nmb
\end{equation}

Eq.(\ref{eq:rhok}) now is simplified to
\begin{equation}
\avg{\rho^k(\epsilon,\phi;L)}=
\bigl(\frac{\rho_0}{2}\bigr)^k
\sum_{m=0}^{\infty}\sum_{\mb=0}^{\infty}
\sum_{m'=0}^{\infty}\sum_{n=0}^{\infty}
e^{i 4 \pi \frac{\phi}{\phi_0}n-\frac{2\eta L}{v(\epsilon)}n}
\overline{Q}\Bigl(\begin{matrix} m,&m',&n\\ \mb,&\mb-m+m',&n \end{matrix} \Bigr|x\Bigr)
\Phi_k(m,\mb,m',\mb-m+m',n,n)
\label{eq:rhok_simplified}
\end{equation}
where
\begin{equation}
\Phi_k(m,\mb,m',\mb',n,\nb)=
\sum_{l=0}^k \sum_{\np=0}^n \sum_{\npb=0}^{\nb} \binom{k}{l}
e^{-i4\pi\frac{\phi}{\phi_0}(\np+\npb)}
\phit_l(m,m';\np,n-\np)
\phit_{k-l}(\mb,\mb';\npb,\nb-\npb)
\label{eq:Phi}
\end{equation}

Expressions (\ref{eq:rhok_simplified}) and (\ref{eq:Phi}) show that
the dominating contribution to $\avg{\rho^k}$ does not strongly
oscillate with the energy. Unlike the averaged \dos, the first
harmonic of all moments oscillates with the halved magnetic flux
$\frac{\phi_0}{2}$.

In the following section, we solve
Eqs.(\ref{eq:comb_transformed})-(\ref{eq:Phi}) for large rings, with
$L \gg \max\{\lp,\lm\}$.

\section{Distribution functions for the weak localization limit}
\label{sec:distribution}

The equation for $\overline{Q}$ is simplified considerably in the limit of large
rings, $L\gg \max\{\lp,\lm\}$.
For this case, we can assume that the electrons are quasi--localized and that the wave function overlaps around the ring are small, similar to a tight--binding model.
Diagrammatically, this means that the electron loops emerging from the l.h.s.\ and the r.h.s.\ of the diagram almost never reach each other, since they have a characteristic size of $\xi \ll L$. (The localization length for an infinite
\oneD\ system is\cite{berezinskii73a,gogolin82} $\xi_{1D} \sim 4 \lm$.)

As a consequence, we can for large rings neglect those inclusions of
the vertices c),c'),e), and f) that directly connect the loops on the 
r.h.s.\ with those on the l.h.s.. Corresponding to this is the neglect
of the terms 2,3, and 6-10 on the r.h.s of Eq.(\ref{eq:Q_DGL}).
Now, $\overline{Q}$ can be factored as

\begin{equation}
\overline{Q}\Bigl(\begin{matrix} m,&m',&n\\ \mb,&\mb-m+m',&n \end{matrix} \Bigr|x\Bigr)
= Q^*(m,\mb,n;x)Q^*(m',\mb-m+m',n;x)
\label{eq:Q_decomposition}
\end{equation}
where the factors are defined through
%This non-entanglement of the loops on the right and left side,
%considered in this approximation means that the terms 2,3, and 6-10 
%on the r.h.s of Eq.(\ref{eq:Q_DGL}), produced by those inclusions of
%vertices c),c'),e), and f) that connect the two sides of the diagram,
%give only a vanishing
%contribution to $Q$. So we can rewrite this equation in terms of $Q^*$
%as
\begin{equation}
\begin{split}
  \frac{\dif Q^*(m,\mb,n;x)}{\dif x}=&
  -\bigl[\frac{2\eta}{v(\epsilon)}(m+\mb)+\frac{2}{\lp}(m-\mb)^2+\frac{1}{\lm}m(m+n)+\frac{1}{\lm}\mb(\mb+\nb)+\frac{1}{\lm}n\bigr] Q^*(m,\mb,n;x)\\
  &+\frac{1}{\lm}m\mb
  Q^*(m-1,\mb-1,n;x)+\frac{1}{\lm}(m+n)(\mb+\nb)Q^*(m+1,\mb+1,n;x)
\end{split}
\label{eq:Qstar_DGL}
\end{equation}

%begin Envers text
Apart from this simplification, the limit $L \gg \max(\lp,\lm)$ implies $m,\mb,m' \gg n$.
%end Envers text

Note that Eq.(\ref{eq:Qstar_DGL}) can also be obtained from Eqs.(\ref{eq:Q_DGL}) and (\ref{eq:Q_substitution})
by neglecting $m'$ and $\mb'$.  The non-entanglement mentioned above has a
second consequence: The remaining contributions change $m$ and $\mb$
simultaneously by $\pm 1$ (due to vertices e) and f) of Fig.\ref{fig:vertices}), or conserve both $m$ and $\mb$, as in Berezinskii's approach to strictly \oneD\ systems.\cite{berezinskii73a} Therefore we can adopt $m=\mb$,
which further simplifies Eq.(\ref{eq:Qstar_DGL}):
\begin{equation}
\begin{split}
  \lm \frac{\dif Q^*(m,n;x)}{\dif x}=&
  -\bigl[4 \eta \tau^- m+n+2m(m+n)\bigr] Q^*(m,n;x) +m^2 Q^*(m-1,n;x)\\
  &+(m+n)^2 Q^*(m+1,n;x)
\end{split}
\label{eq:Qstar_DGL_simplified}
\end{equation}
Here, $\tau^-$ is the inelastic scattering time with respect to
backward scattering.  The boundary condition for
Eq.(\ref{eq:Qstar_DGL_simplified}) is
\begin{equation}
Q^*(m,n;x=0)=\delta_{m,0}
\label{eq:Qstar_BC}
\end{equation}

The combined mixing function $\Phi_k$, given in Eq.(\ref{eq:Phi}) is
simplified for large $m$, $\mb$, and $m'$ in Appendix~\ref{app:Phi_simplify}.
Substituting Eqs.(\ref{eq:Phi_sol}) and (\ref{eq:Q_decomposition}) into (\ref{eq:rhok_simplified}), we get a comparatively simple expression for $\avg{\rho^k}$:
\begin{equation}
\begin{split}
\avg{\rho^k(\epsilon,\phi;L)}=
&\bigl(\frac{\rho_0}{2}\bigr)^k
\sum_{m=0}^{\infty}\sum_{m'=0}^{\infty}
\sum_{n=0}^{\infty}\sum_{l=0}^k
\bigl[\cos(2\pi \frac{\phi}{\phi_0})\bigr]^{2n}
e^{-\frac{2\eta L}{v(\epsilon)}n}
\binom{k}{l} \binom{2l}{m-m'} \binom{2k-2l}{m-m'}\\
& \frac{2^{2n}m^{k+2n-2}}{\Gamma(l)\Gamma(k-l)}
Q^*(m,n;L)Q^*(m',n;L)
\end{split}
\label{eq:rhok_simplified1}
\end{equation}
As we emphasized, the diagrammatical structure of the block $Q^*$ demands its dependence on one parameter $m$ instead of two ($m$ and $\mb$). Hereby the sum over $\mb$ is removed. The summations over $l$ and $m'$ can be done as described in Appendix~\ref{app:Phi_simplify}. Using Eqs. (\ref{eq:rhok_simplified2}) and (\ref{eq:wieder_ein_label}), we get
\begin{equation}
\begin{split}
\avg{\rho^k(\epsilon,\phi;L)}&=
\rho_0^k
\frac{2^{1-k}(k-1)}{k(2k-1)}
\frac{\Gamma^2(2k)}{\Gamma^5(k)}
\sum_{n=0}^{\infty} \sum_{m=0}^{\infty}
\frac{2^{2n} e^{-\frac{2\eta L}{v(\epsilon)}n} m^{k+2n-2}}{\Gamma^2(n+1)}
\bigl[\cos(2\pi \frac{\phi}{\phi_0})\bigr]^{2n}
{Q^*}^2(m,n;L)\\
&=\avg{\rho^k(\epsilon,\phi;L)}_{n=0}+
\avg{\rho^k(\epsilon,\phi;L)}_{n=1} \cos^2(\frac{2\pi\phi}{\phi_0})+\dots
\end{split}
\label{eq:rhok_final}
\end{equation}
Eq.(\ref{eq:Qstar_DGL_simplified}) for $Q^*(m,n;x)$ was solved approximately in [\onlinecite{nakhmedov90a}] for arbitrary $n$. Here we shall study this equation for the zeroth and first harmonics ($n=0$ and $n=1$) in detail.

\subsection{Zeroth harmonic contribution to the \dos\ moments}

The zeroth harmonic of $\avg{\rho^k(\epsilon,\phi;L)}$ in Eq.(\ref{eq:rhok_final}) contains 
$Q^*(m,n=0;L)$.
Laplace transforming Eq.(\ref{eq:Qstar_DGL_simplified}), written for $n=0$, with respect to $x$ and using the boundary condition (\ref{eq:Qstar_BC}), we get
\begin{equation}
(\lambda + s_1 m)Q^*_0(m;\lambda)-\delta_{m,0}=m^2\bigl[ Q^*_0(m+1;\lambda) + Q_0^*(m-1;\lambda)-2Q_0^*(m;\lambda)\bigr]
\label{eq:Qstarlaplace_equation}
\end{equation}
Here, $s_1= 4 \eta \tau^-$ and $\lambda$ is the parameter of the Laplace transform. This is an equation for the right hand side in the Berezinskii technique with an open boundary condition and it was solved in [\onlinecite{melnikov80a,altshuler89a}]. Here, we give only the result for $Q_0^*(s_1,m;x)$:
\begin{equation}
Q_0^*(s_1,m;x)= 2 (m s_1)^\frac{1}{2} {\rm K}_1(2(m s_1)^\frac12)
+\frac{2  (m s_1)^\frac{1}{2}}{2 \pi i} \int_{-\infty}^{\infty}\dif \lambda
\frac{s_1^{-\frac{1+i \lambda}{2}}}{i-\lambda}
\frac{\Gamma^3(\frac{1-i\lambda}{2})}{\Gamma^2(-i\lambda)}
e^{-(\lambda^2+1)\frac{x}{4 \lm}}
{\rm K}_{-i\lambda}(2(m s_1)^\frac{1}{2})
\label{eq:Qstar0_sol}
\end{equation}
After substitution of this solution into $\avg{\rho^k(\epsilon,\phi;L)}_{n=0}$ in Eq.(\ref{eq:rhok_final}), the summation over $m$ can be transformed into an integration, which is done easier. Some mathematics results
in the following form of $\avg{\rho^k(\epsilon,\phi;L)}_{n=0}$:
\begin{equation}
\begin{split}
\avg{\rho^k(\epsilon,\phi;L)}_{n=0}=&\bigl(\frac{\rho_0}{2}\bigr)^k
\frac{2(k-1)\Gamma(2k)}{k(2k-1)\Gamma^5(k)}s_1^{1-k}
\Bigl\{\frac{k}{k-1}\Gamma^4(k) \\
&+ \frac{2}{\sqrt{s_1}}e^{-\frac{L}{4}}
\int_{-\infty}^{\infty}\frac{\dif \lambda}{2 \pi i}
\frac{e^{-\frac{L}{4 \lm}(\lambda-i\gamma)^2 - \frac{L}{4\lm}\gamma^2}}{i-\lambda}
\frac{\Gamma^3\bigl(\frac{1-i\lambda}{2}\bigr)}{\Gamma^2(-i\lambda)}
\Bigl|\Gamma\bigl(\frac{2k+1+i\lambda}{2}\bigr)
\Gamma\bigl(\frac{2k-1+i\lambda}{2}\bigr)\Bigr|^2\\
&+\frac{2}{s_1}\int_{-\infty}^{\infty}\frac{\dif z'}{2\pi i}
e^{-\frac{L}{2\lm} {z'}^2}
|\Gamma(k-i z')|^2\\
&\int_{-\infty}^{\infty}\frac{\dif z}{2\pi i}
\frac{e^{-\frac{L}{2\lm}(z-i\gamma)^2-\frac{L}{2\lm}\gamma^2-\frac{L}{2\lm}}}
{(z+z'-i)(z-z'-i)}
\frac{\Gamma^3\bigl(\frac{1-iz-iz'}{2}\bigr)
\Gamma^3\bigl(\frac{1-iz+iz'}{2}\bigr)}
{\Gamma^2(-iz-iz')\Gamma^2(-iz+iz')}
|\Gamma(k-iz)|^2\Bigr\}
\end{split}
\label{eq:rhok0_1}
\end{equation}
where $\gamma=\frac{\lm}{L}\ln\frac{1}{s_1}>0$.  The second term in the
bracket of Eq.(\ref{eq:rhok0_1}) has a saddle point at
$\lambda_0=i\gamma$ and simple poles at the upper half--plane:
$\lambda_1=i,\lambda_2=i(2k-1),i(2k+1),\dots$. The integral over $z$
in the third term in the bracket contains again the saddle point at
$z_0=i\gamma$ and poles at $z_1=\pm z'+i,z_2=ik,i(k+1),\dots$.  For
$\gamma < 1$, the main contribution to both integrals is given by the
saddle points. As a result we get
\begin{equation}
\avg{\rho^k(\epsilon,\phi;L)}_{n=0}=\rho_0^k(2s_1)^{1-k}\frac{\Gamma(2k-1)}{\Gamma(k)}
\label{eq:rhok0_2}
\end{equation}
Such a result has been obtained for the infinite \oneD\ disordered
system.\cite{altshuler89a} Transforming the semi--invariants in
Eq.(\ref{eq:rhok0_2}) to moments and using the inverse Mellin
transformation
\begin{equation}
W(\rho)=\frac{1}{2\pi i} \int_{a - i \infty}^{a + i \infty}
\frac{\dif k}{\rho^{k+1}} \langle \rho^k \rangle
\label{eq:inverse_Mellin}
\end{equation}
the following inverse--Gaussian distribution function is obtained:
\begin{equation}
W_{n=0}(\rho)=\bigl(\frac{2 \eta \tau^- \rho_0}{\pi \rho^3}\bigr)^\frac{1}{2}
\exp\bigl(-2 \frac{(\rho-\rho_0)^2}{\rho \rho_0} \eta \tau^-\bigr)
\end{equation}
For $2 \eta \tau^- \gg 1$, the most probable or typical value of
$\rho$ is equal to $\rho_0$, whereas for $2 \eta \tau^- \ll 1$ it
shifts to lower values and becomes equal to $\rho_{\rm typ}= \frac{4
  \eta \tau^-}{3} \rho_0$.

When $\gamma$ assumes intermediate values, i.e.\ $1<\gamma<k$, the
essential contribution to $\avg{\rho^k}_{n=0}$ comes from the saddle
points of the third term in the bracket of Eq.(\ref{eq:rhok0_1}) and
the contributions from the poles of this term cancel the other term,
resulting in
\begin{equation}
\avg{\rho^k(\epsilon,\phi;L)}_{n=0}=
\frac{\rho_0^k \Gamma(2k-1)}{2^{k-1}\Gamma(k-1)\Gamma(k+1)}
\frac{s_1^{-k} \lm e^{-\frac{L}{2\lm}(1+\gamma^2)}}{\pi L (1-\gamma)^2}
\frac{\Gamma^6\bigl(\frac{1+\gamma}{2}\bigr)}{\Gamma^4(\gamma)}
\Gamma(k+\gamma)\Gamma(k-\gamma)
\label{eq:rhok0_3}
\end{equation}
This expression shows that high moments of the \dos\ for intermediate
values of $\gamma$ increase with $k$; however the increase is not so
rapid, Eq.(\ref{eq:rhok0_3}) has an additional factor
$\frac{1}{k!}$ compared to Eq.(\ref{eq:rhok0_2}).

For $\gamma$ satisfying the condition $\gamma > k$, the leading
contribution is given by the pole $z_2= ik$ and
\begin{equation}
\avg{\rho^k(\epsilon,\phi;L)}_{n=0} = \bigl(\frac{\rho_0}{2}\bigr)^k
\sqrt{\frac{\lm}{2\pi L}}
\frac{4}{k(k-1)(2k-1)}
\frac{\Gamma^2(2k)}{\Gamma^7(k)}\Gamma^6\bigl(\frac{k+1}{2}\bigr)
e^{\frac{L}{2\lm}(k^2-1)}
\label{eq:rhok_gamma_smaller_k}
\end{equation}
The last expression is valid for arbitrary small values of the
dissipation parameter ($\eta \to 0$ or $\gamma \to \infty$) with
$\eta \ll \frac{1}{4\tau^-}\exp(\frac{-k L}{\lm})$.
Eq.(\ref{eq:rhok_gamma_smaller_k}) shows that the zeroth harmonic of the $k$th
moment of the \dos\ grows with $k$ as $\exp(k^2)$. Such rapid
increasing of high moments of $\avg{\rho^k}_{n=0}$ has been firstly obtained
by Wegner\cite{wegner80b} and it is a characteristic
feature of the logarithmic normal distribution of
$\avg{\rho^k}_{n=0}$. The distribution function for the zeroth
harmonic term can be obtained using Eq.(\ref{eq:inverse_Mellin}).  For
large values of the \dos, satisfying the condition $\rho >
\frac{\rho_0}{2}\exp(\frac{L}{\lm})$, the dominating saddle point
yields again a logarithmic normal distribution:
\begin{equation}
W_{n=0}(\rho)=\frac{8\lm}{\pi \rho_0 L}
\frac{\Gamma\bigl(\frac{2\lm}{L}\ln\frac{2\rho}{\rho_0}-2\bigr)
\Gamma\bigl(\frac{2\lm}{L}\ln\frac{2\rho}{\rho_0}\bigr)
\Gamma^6\bigl(\frac{1}{2}+\frac{\lm}{2 L}\ln\frac{2\rho}{\rho_0}\bigr)}
{\Gamma\bigl(\frac{\lm}{L}\ln\frac{2\rho}{\rho_0}+1\bigr)
\Gamma^6\bigl(\frac{\lm}{L}\ln\frac{2\rho}{\rho_0}\bigr)}
\exp\Bigl[-\frac{\lm}{2L}\bigl(\ln\frac{2\rho}{\rho_0}+\frac{L}{\lm}\bigr)^2\Bigr]
\label{eq:W0}
\end{equation}
For small values of $\rho$, when $\rho<\frac{\rho_0}{2}\exp(\frac{L}{\lm})$,
the main contribution comes from the pole at the origin and the distribution
function decreases power--like:
\begin{equation}
W_{n=0}(\rho)=\frac{2\rho_0}{\rho^2}\sqrt{\frac{\lm}{2 \pi L}}
\end{equation}
Thus the distribution function for the zeroth harmonic or $\phi$-independent component has asymmetric form.
 
\subsection{Amplitude of the first harmonic contribution to the \dos\ moments}

By a Laplace transform with respect to $x$, Eq.(\ref{eq:Qstar_DGL_simplified}) with $n=1$ is converted to
\begin{equation}
(\lambda+ s_1m)Q_1^*(m;\lambda)-\delta_{m,0}=(m+1)^2[Q_1^*(m+1;\lambda)-Q_1^*(m;\lambda)]+m^2[Q_1^*(m-1;\lambda)-Q_1^*(m;\lambda)]
\label{eq:Qstar1_laplace}
\end{equation}
The $\delta$ symbol on the left--hand side of this equation comes from the boundary condition (\ref{eq:Qstar_BC}).
Eq.(\ref{eq:Qstar1_laplace}) corresponds to the equation for the central part in the Berezinskii technique for strictly \oneD\ systems with open boundary.\cite{melnikov80a} For $m \gg 1$, this equation is transformed into a differential equation,
\begin{equation}
m^2\frac{\dif^2 Q_1^*(m;\lambda)}{\dif m^2}+2m\frac{\dif Q_1^*(m;\lambda)}{\dif m}-(\lambda+s_1 m)Q_1^*(m;\lambda)=0
\label{eq:Qstar1_DGL_in_m}
\end{equation}
A change of the function to $\frac{1}{z} \Phi(z,\lambda)=Q_1^*(m;\lambda)$, 
where $z^2=4 m s_1$, reduces Eq.(\ref{eq:Qstar1_DGL_in_m}) to the Bessel equation
\begin{equation}
\frac{\dif^2 \Phi}{\dif z^2} + \frac{1}{z} \frac{\dif \Phi}{\dif z} - \bigl(1+\frac{1+4\lambda}{z^2}\bigr) \Phi=0
\end{equation}
Therefore $Q_1^*(m;\lambda)$ can be expressed as
\begin{equation}
Q_1^*(m;\lambda) = C \frac{1}{2 (m s_1)^\frac12} {\rm K}_{1+2q}(2 (m s_1)^\frac12)
\label{eq:Qstar1_bessel}
\end{equation}
where $q=-\frac12 + \sqrt{\lambda + \frac14}$.

Eq.(\ref{eq:Qstar1_bessel}) contains an unknown parameter $C$ due to the neglect of the Kronecker symbol in Eq.(\ref{eq:Qstar1_laplace}).

On the other hand Eq.(\ref{eq:Qstar1_laplace}) has been solved by Mel'nikov,\cite{melnikov80a} who obtained the asymptotic solution of $Q_1^*(m;\lambda)$ for $1\ll m \ll s_1^{-1}$ as
\begin{equation}
Q_1^*(m;\lambda)=\frac{\Gamma^3(q+1)}{\Gamma(2q+2)} m^{-q-1}
\label{eq:Qstar1_small_m}
\end{equation}
The comparison or Eq.(\ref{eq:Qstar1_bessel}) with the asymptotic form (\ref{eq:Qstar1_small_m})
allows to determine $C$:
\begin{equation}
C=4 s_1^{q+1}\frac{\Gamma^3(q+1)}{\Gamma(2q+2)\Gamma(2q+1)}
\label{eq:C}
\end{equation}
Taking the inverse Laplace transform, one obtains for $Q_1^*(m;x)$
\begin{equation}
Q_1^*(m;x)=\int_{-\infty}^{\infty}\frac{\dif \lambda}{2 \pi} 
e^{-\frac14 (\lambda^2+1) \frac{x}{\lm}} s_1^{\frac{1-i \lambda}{2}}
(m s_1)^{-\frac12} {\rm K}_{-i \lambda}(2 (m s_1)^\frac12)
\frac{\Gamma^3\bigl(\frac{1-i\lambda}{2}\bigr)}{\Gamma^2(-i\lambda)}
\label{eq:Qstar1_transformed_back}
\end{equation}
To get an expression for the first harmonic, $\avg{\rho^k(\epsilon,L)}_{n=1}$, we substitute the solution (\ref{eq:Qstar1_transformed_back}) into Eq.(\ref{eq:rhok_final}), and sum over $m$, which can be done after the transformation of
the sum into an integral over $\kappa=m s_1$:
\begin{equation}
\begin{split}
\avg{\rho^k(\epsilon,L)}_{n=1}=& \bigl(\frac{\rho_0}{2}\bigr)^k
\frac{(k-1)\Gamma(2k)}{k \Gamma^5(k)} s_1^{-k} e^{-\frac{L}{2\lm}}
\int_{-\infty}^{\infty} \frac{\dif \lambda}{2\pi}
s_1^{-\frac{i\lambda}{2}} e^{-\frac{L}{4\lm}\lambda^2}
\frac{\Gamma^3\bigl(\frac{1-i\lambda}{2}\bigr)}{\Gamma^2(-i\lambda)}\\
&\int_{-\infty}^{\infty}\frac{\dif \lambda'}{2\pi} 
s_1^{-\frac{i\lambda'}{2}} e^{-\frac{L}{4\lm}{\lambda'}^2}
\frac{\Gamma^3\bigl(\frac{1-i\lambda'}{2}\bigr)}{\Gamma^2(-i\lambda')}
\Bigl|\Gamma\bigl(k+\frac{i\lambda + i \lambda'-1}{2}\bigr)
\Gamma\bigl(k+\frac{i\lambda - i \lambda'-1}{2}\bigr)\Bigr|^2
\end{split}
\label{eq:Qstar1_sol}
\end{equation}

For convenience, we substitute below the variables $\lambda$ and $\lambda'$ 
by $z$ and $z'$ according to $\lambda=z+z'$ and $\lambda'=z-z'$.

The values of the integrals in Eq.(\ref{eq:Qstar1_sol}) are determined by saddle
points and poles. For $\gamma=\frac{\lm}{L}\ln\frac{1}{s_1}<k-\frac12$, 
the contribution from the saddle point dominates:
\begin{equation}
\avg{\rho^k(\epsilon,L)}_{n=1}=\bigl(\frac{\rho_0}{2}\bigr)^k
\frac{(k-1)\lm \Gamma^2(2k) s_1^{-k}}{k\pi L \Gamma^5(k)}
e^{-\frac{L}{2\lm}(1+\gamma^2)}
\frac{\Gamma^2\bigl(\frac{2k-1}{2}\bigr)
\Gamma^6\bigl(\frac{1+\gamma}{2}\bigr)
\Gamma\bigl(\frac{2k-1+2\gamma}{2}\bigr)
\Gamma\bigl(\frac{2k-1-2\gamma}{2}\bigr)}
{\Gamma^4(\gamma)}
\end{equation}
For $\gamma > k-\frac12$ the main contribution is given by the pole at 
$z=i(k-\frac12)$ and one gets
\begin{equation}
\avg{\rho^k(\epsilon,L)}_{n=1}=\bigl(\frac{\rho_0}{2}\bigr)^k
\frac{2(k-1)\Gamma(2k)\Gamma(2k-1)\Gamma^6\bigl(\frac{2k+1}{4}\bigr)}
{k \Gamma^5(k)\Gamma^2\bigl(\frac{2k-1}{2}\bigr)}
\sqrt{\frac{\lm}{2\pi L s_1}}
e^{\frac{L}{8\lm}(2k-1)^2-\frac{L}{2\lm}}
\label{eq:rhok1}
\end{equation}
In contrast to the expression of $\avg{\rho^k(\epsilon,L)}_{n=0}$ for small 
dissipation, $\eta \to 0$ [see Eq.(\ref{eq:rhok_gamma_smaller_k})], the expression for 
$\avg{\rho^k(\epsilon,L)}_{n=1}$ increases strongly with $s_1 = 4 \tau^-\eta \to 0$. It is illustrative to rewrite the prefactor of Eq.(\ref{eq:rhok1}) as
$(2 \pi \frac{L}{\lm}s_1)^\frac12 = ( 8 \pi^2 \frac{\eta}{\Delta})^\frac12$
in terms of the level distance $\Delta = \frac{1}{\rho_0 L}$ and the 
dissipation energy $\eta$, the latter blurring the quantized energy levels.
By decreasing $\eta$, the energy levels are sharpened and the distribution
function becomes a $\delta$--function.

Substituting Eq.(\ref{eq:rhok1}) into (\ref{eq:inverse_Mellin}), one receives 
a normal logarithmic distribution for $\rho > \frac{\rho_0}{2}\exp(-\frac{L}{\lm})$:
\begin{equation}
W_{n=1}(\rho)=\frac{\lm}{\pi L}
\sqrt{\frac{2}{s_1\rho\rho_0}}
\frac{\Gamma\bigl(1+\frac{2\lm}{L}\ln\frac{2\rho}{\rho_0}\bigr)
\Gamma\bigl(\frac{2\lm}{L}\ln\frac{2\rho}{\rho_0}\bigr)
\Gamma^6\bigl(\frac{1}{2}+\frac{\lm}{2 L}\ln\frac{2\rho}{\rho_0}\bigr)}
{\Gamma\bigl(\frac32 + \frac{\lm}{L}\ln\frac{2\rho}{\rho_0}\bigr)
\Gamma\bigl(\frac{\lm}{L}\ln\frac{2\rho}{\rho_0}-\frac12\bigr)
\Gamma^3\bigl(\frac{\lm}{L}\ln\frac{2\rho}{\rho_0}+\frac12\bigr)
\Gamma^2\bigl(\frac{\lm}{L}\ln\frac{2\rho}{\rho_0}\bigr)}
\exp\Bigl[-\frac{\lm}{2L}\bigl(\ln\frac{2\rho}
{\rho_0}+\frac{L}{\lm}\bigr)^2\Bigr]
\label{eq:W1}
\end{equation}
The logarithmic normal distribution function for the first harmonic is
valid for a large range of $\rho$. Comparing Eq.(\ref{eq:W1}) with
Eq.(\ref{eq:W0}) for winding number zero, it can be seen that
Eq.(\ref{eq:W1}) contains in addition a prefactor
$\sqrt{\frac{\rho_0}{\eta \tau^-\rho}}= (\pi \eta \lm \rho)^{-1/2}$
which increases with decreasing temperature. Thus the first harmonic
increases with decreasing temperature faster in amplitude than the
zeroth harmonic.

\section{Conclusion}
\label{sec:conclusion}
The distribution function for the local \dos\ in a one--channel ring
threaded by a magnetic flux through the opening was studied in this
paper.  For this purpose, we constructed a new diagrammatic method as
an extension of the Berezinskii technique\cite{berezinskii73a} to the
problem with periodic boundary conditions and in the presence of an
external magnetic field. The equations obtained ((\ref{eq:rho}) to
(\ref{eq:BC_for_Q0}) and (\ref{eq:rhok}) to (\ref{eq:Q_BC}) for the
\dos\ and its $k$th moments, respectively) are exact in the framework of
the weak disorder limit $k_F l \gg 1$.  Eqs.(\ref{eq:DGL_for_Q0}) and
(\ref{eq:BC_for_Q0}) are solved exactly, which gives the oscillation
of $\rho$ with the full flux for both weak localization and
ballistic regimes.

In contrast to the \dos\ problem, the equation for
$\avg{\rho^k(\epsilon,\phi;L)}$ is rather complicated and we succeeded
to solve it for the weak localization limit when $L\gg l^\pm$. In this
limit, the leading contributions to arbitrary moments of the \dos\
oscillate with the halved period $\frac{\phi_0}{2}$. The distribution
functions for zeroth (insensitive to the magnetic field) and first
(with a period of $\frac{\phi_0}{2}$) harmonics are calculated and
logarithmic normal distributions (Eqs.(\ref{eq:W0}) and (\ref{eq:W1})) are
obtained for them, indicating large contributions from high
  moments of the \dos. For the zeroth harmonic, this normal logarithmic
shape appears for the tail of the distribution, but for the first harmonic it
covers the large range of $\rho >
\frac{\rho_0}{2}\exp(-\frac{L}{\lm})$, i.e.\ the high moments give
essential contributions not only on the tail but also in the vicinity
of the average value of the \dos.  The distribution function for the
first harmonic increases with decreasing the width of the energy
levels or the dissipation parameter $\eta$ (see Eq.(\ref{eq:W1})), which was
introduced phenomenologically in the theory (Eq.(\ref{eq:bareGF})).
For $\eta \to 0$, the distribution function $W_{n=1}(\rho)$ becomes a
$\delta$-function due to the quantization of the energy levels in the
rings.  The results for the \dos\ show that the amplitudes of all
harmonics of $\rho(\epsilon, \phi)$ are exponentially small in the
weak localization regime [Eq.(\ref{eq:rho_weak_localization})], while
the amplitudes of the higher moments in this regime [Eqs.(\ref{eq:rhok0_2}),
(\ref{eq:rhok0_3}), (\ref{eq:rhok_gamma_smaller_k}), and (\ref{eq:rhok1})] are relatively large.
Although we could not calculate higher moments of the \dos\ in the
ballistic regime, the amplitude of the average value of the \dos\ is
large and seems to be consistent with experimental
data.\cite{mailly93a}

It is also well known that the \dos\ of \oneD\
disordered crystalline systems is very sensitive to the filling
factor. There exists disorder induced enhancement of the \dos\ for
commensurable values of the electron wavelength $\lambda$ and the
lattice constant $a$, when the electronic energy $\epsilon$ satisfies
the condition $p(\epsilon)=\frac{k \pi}{n a}$, $k= \pm 1, \pm 2, \dots
\pm n$ and $n=2,3,\dots$, and the effect is pronounced for
half--filling which corresponds to $n=2$. The singularity in the \dos\
of \oneD\ disordered crystalline systems near the middle of the band is
known as a Dyson singularity\cite{dyson53a} which was studied for many
\oneD\ electronic models.\cite{weissman75a,gorkov76a,hirsch76a,ovchinnikov77a,ovchinnikov77b,gogolin77a}
Notice that the Berezinskii method has also been
applied to study the conductivity and the localization
length\cite{gogolin77a,gogolin82} apart from the Dyson singularity in
the middle of the band of a \oneD\ infinite lattice with both weak and
strong disorder.\cite{gogolin77a,gogolin79a}
Our preliminary study shows that the
real space diagrammatic method presented in this paper is applicable to
study the Dyson singularity in the \dos\ of a ring for a half--filled energy
band.
This leads to a remarkable high amplitude of the persistent current as it is observed in the experiments, provided that the Peierls transition is suppressed by impurities and by weak transvers tunneling between the channels.

The authors thank M. Kiselev for discussion.
This work was supported by the SFB410.

%Appendix Appendix Appendix Appendix Appendix Appendix Appendix Appendix
%Appendix Appendix Appendix Appendix Appendix Appendix Appendix Appendix
\appendix
\section{Solution for $Q_0(m,n;x)$}
\label{app:Calc_of_Q0}

The first term on the right hand side of Eq.(\ref{eq:DGL_for_Q0}) can
be removed through the transformation
\begin{equation}
Q_0(m,n;x)=\exp\bigl\{-\frac{x}{2\lp}(2m+n)^2-\frac{x}{2\lm}n-\frac{x}{\lm}m(m+n)\bigr\}Q_0^*(m,n;x)
\label{eq:Q0_transform1}
\end{equation}
which gives for Eqs.(\ref{eq:DGL_for_Q0}) and (\ref{eq:BC_for_Q0}) the
simpler form
\begin{equation}
\lm \frac{\dif Q_0^*(m,n;x)}{\dif x}=-m^2 \exp\bigl\{\frac{xn}{\lm}\bigr\}Q_0^*(m-1,n+2;x)
\quad\mbox{with}\quad
Q_0^*(m,n;x=0)=\delta_{m,0}
\end{equation}
Laplace transformation from $x$ to $\lambda$ yields
\begin{equation}
\lambda \overline{Q_0}(m,n;\lambda) - \delta_{m,0}=-\frac{1}{\lm}m^2\overline{Q_0}(m-1,n+2;x-\frac{n}{\lm})
\label{eq:Q0laplace_equation}
\end{equation}
Eq.(\ref{eq:Q0laplace_equation}) can be solved by iteration:
\begin{equation}
\begin{split}
  \overline{Q_0}(m,n;\lambda)=\bigl(\frac{-1}{\lm}\bigr)^m (m!)^2 \prod_{j=0}^m \frac{1}{\lambda-\frac{1}{\lm}j(j+n-1)}\\
  =\frac{(m!)^2}{\lambda}
  \frac{\Gamma(z+1+\frac{n-1}{2})\Gamma(\frac{n-1}{2}+1-z)}{\Gamma(\frac{n-1}{2}+m+1+z)\Gamma(\frac{n-1}{2}+m+1-z)}
\end{split}
\label{eq:Q0laplace_sol}
\end{equation}
where $z^2=\lambda \lm + \frac{(n-1)^2}{4}$. The inverse Laplace
transform gives for $Q_0^*$
\begin{equation}
Q_0^*(m,n;x)=(m!)^2\sum_{j=0}^m \exp\bigl\{\frac{x}{\lm}j(j+n-1)\bigr\} \frac{(-1)^j}{j!(m-j)!}
\frac{(j+n-2)!}{(m+j+n-1)!}(2j+n-1)
\label{eq:Q0star_sol}
\end{equation}
which, in connection with (\ref{eq:Q0_transform1}), gives the final
result Eq.(\ref{eq:Q0_sol}), where $x$ is replaced by the full
circumference $L$. The compliance of Eq.(\ref{eq:Q0star_sol}) with the
boundary condition is easily checked. Also, for $m=0$,
$Q_0^*(0,n;x)=1$, and for $n=0$ we get from the inverse Laplace
transform of Eq.(\ref{eq:Q0laplace_sol}) or from taking the limit of
Eq.(\ref{eq:Q0star_sol})
\begin{equation}
Q_0^*(m,0;x)=(1-m-m\frac{x}{\lm})+\sum_{j=2}^m \exp\bigl\{\frac{x}{\lm}j(j+n-1)\bigr\}\binom{m}{j}(-1)^j\frac{m! (j+n-2)!}{(m+j+n-1)!}(2j+n-1)
\label{eq:Q0star_n0}
\end{equation}

\section{Calculation of the mixing coefficient}
\label{app:Phi_simplify}
Using the relations
\begin{equation}
\delta_{m,k}=\oint_{|z|<1}\frac{\dif z}{2 \pi i} z^{k-m-1}
\quad\mbox{and}\quad
\binom{m}{k}=\oint_{|z|<1}\frac{\dif z}{2 \pi i} \frac{1}{z^{k+1}(1-z)^{m-k+1}}
\end{equation}
we can transform Eq.(\ref{eq:comb_transformed}) for
$\phit_l(m,m';\np,n-\np)$ to
\begin{equation}
\phit_l(m,m;\np,n-\np)=
\oint\frac{\dif z_1}{2 \pi i} \frac{1}{z_1^{m+1}}
\oint\frac{\dif z_2}{2 \pi i} \frac{1}{z_2^{m'+1}}
\oint\frac{\dif z_3}{2 \pi i} \frac{1}{z_3^{\np+1}}
\oint\frac{\dif z_4}{2 \pi i} \frac{1}{z_4^{n-\np+1}}
\Bigl\{\frac
{(1+z_1)(1+z_2)-z_3z_4}
{(1-z_3)(1-z_4)-z_1z_2}\Bigr\}^l
\label{eq:phit1}
\end{equation}
Substituting $z=z_1z_2$, the dominant contribution for $m\gg1$ comes
from the pole at $z=(1-z_3)(1-z_4)$. Integrating over this new variable
gives
\begin{equation}
\begin{split}
  \phit_l(m,m';\np,n-\np)=&\frac{(m+l-1)!}{(l-1)!m!}  \oint\frac{\dif
    z_2}{2 \pi i}\frac{1}{z_2^{m'-m+l+1}} \oint\frac{\dif z_3}{2 \pi
    i} \frac{1}{z_3^{\np+1}}
  \oint\frac{\dif z_4}{2 \pi i} \frac{1}{z_4^{n-\np+1}}\\
&  \frac{\bigl[(z_2+(1-z_3)(1-z_4))(1+z_2)-z_2z_3z_4\bigr]^l}
  {(1-z_3)^{m+l}(1-z_4)^{m+l}}
\label{eq:phit2}
\end{split}
\end{equation}
The remaining integrals are done in a similar way, resulting in

\begin{equation}
\phit_l(m,m';\np,n-\np)=\frac
{(2l)!(m+l+\np-1)!(m+l+n-\np-1)!}
{(m'-m+l)!(m-m'+l)!(m+l-1)!\np!(n-\np)!(l-1)!m!}
\label{eq:phit3}
\end{equation}

Now we can collect all $\np$ and $\npb$ dependent terms in
Eq.(\ref{eq:rhok}) and sum over $\np$ and $\npb$, introducing the
mixing function $\Phi_k$ from Eq.(\ref{eq:Phi}).  For large $m$, we
can use Stirlings formula
\begin{equation}
\lim_{m\to\infty} \frac{(m+a)!}{(m+b)!} m^{b-a}=1 
\end{equation}
to obtain
%\begin{equation}
%\begin{split}
%  \Phi_k&(m,\mb,m',\mb',n,\nb)=
%  \sum_{l=0}^k \bigl(1+e^{-4i\pi\frac{\phi}{\phi_0}}\bigr)^{n+\nb}\\
%  &\frac{m^{n+l-1}\mb^{\nb+k-l-1} k! (2l)!(2k-2l)!}
%  {l!(k-l)!(m'-m+l)!(m-m'+l)!(\mb'-\mb+k-l)!(\mb-\mb'+k-l)!(l-1)!(k-l-1)!n!\nb!}
%\end{split}
%\label{eq:Phi_sol} %B6
%\end{equation}

\begin{equation}
\Phi_k(m,\mb,m',\mb',n,\nb)=
\sum_{l=0}^k \frac{\bigl(1+e^{-4i\pi\frac{\phi}{\phi_0}}\bigr)^{n+\nb}}{(l-1)!(k-l-1)!n!\nb!}
\binom{k}{l} \binom{2l}{m-m'+l} \binom{2k-2l}{\mb-\mb'+l}
\label{eq:Phi_sol} %B6
\end{equation}

Taking into account $Q^*(m,\mb,n;x)=Q^*(m,n;x)\delta_{m,\mb}$, this gives Eq.(\ref{eq:rhok_simplified1}).
For $m,m'\gg 1$, the summations over $l$ and $m'$ can be done. Following the paper [\onlinecite{altshuler89a}], we denote $\Delta m=m-m'$, with $\Delta m \ll m$ for large $m$. The significant contributions to Eq.(\ref{eq:rhok_simplified1}) come from $Q^*(m-\Delta m,n;x) \approx Q^*(m,n;x)$. Hence we can rewrite Eq.(\ref{eq:rhok_simplified1}) as
\begin{equation}
\begin{split}
\avg{\rho^k(\epsilon,\phi;L)}=&
\bigl(\frac{\rho_0}{2}\bigr)^k
\sum_{m=0}^{\infty}
\sum_{n=0}^{\infty}
\bigl[\cos(2\pi \frac{\phi}{\phi_0})\bigr]^{2n}
e^{-\frac{2\eta L}{v(\epsilon)}n}
\frac{2^{2n}m^{k+2n-2}k!}
{(n!)^2}
{Q^*}^2(m,n;x)\\
&\sum_{l=0}^k \binom{k}{l} \sum_{\Delta m=-k}^{k}
\binom{2l}{l+\Delta m} \binom{2k-2l}{k-l+\Delta m}
\frac{1}{(l-1)!(k-l-1)!}
\end{split}
\label{eq:rhok_simplified2}
\end{equation}
The last two sums result in
\begin{equation}
\frac{2(k-1)}{k(2k-1)}
\frac{\Gamma^2(2k)}{\Gamma^5(k)}
\label{eq:wieder_ein_label}
\end{equation}
which is used in the final result for the $k$th moment of the \dos, Eq.(\ref{eq:rhok_final}). 

% for submission:
%\input{letter.bbl} 
% for use of bib-Tex:
\bibliographystyle{prsty} \bibliography{oppgroup}

\end{document}